\documentclass[useAMS,usenatbib]{mn2e}
\usepackage{psfig, epsf, epsfig}
\title[Intracluster globulars]{On spatial distributions 
of old globular clusters in clusters of galaxies}
\author[K. Bekki and H. Yahagi]
{Kenji Bekki${}^1$\thanks{E-mail:
bekki@bat.phys.unsw.edu.au} 
and Hideki Yahagi${}^2$\\
       ${}^1$School of Physics, University of New South Wales,
              Sydney 2052, NSW, Australia\\
       ${}^2$ Department of Astronomy, University of Tokyo, 
              7-3-1 Hongo, Bunkyo ward, Tokyo 113-0033, Japan\\}
  
\begin{document}

\date{Accepted, Received 2005 February 20; in original form }

\pagerange{\pageref{firstpage}--\pageref{lastpage}} \pubyear{2005}

\maketitle

\label{firstpage}

\begin{abstract}

We investigate 
structural properties of old, metal-poor
globular clusters (GCs) formed at high redshifts ($z>6$) and located
inside and outside virialized galaxy-scale halos in clusters
of galaxies with the total masses of $M_{\rm CL}$
based on
high-resolution cosmological
simulations with models of GC formation. 
We mainly derive the parameter dependences of physical properties
of intracluster GCs (ICGCs) based on the results of 14 cluster models.
Our principle results 
are summarized as follows.

(1) ICGCs are formed as a result of tidal stripping of GCs initially
within galaxy-scale halos
during hierarchical growth of clusters via halo merging.
These ICGCs comprise 20-40 \% of all GCs
in clusters with 
$1.0 \times 10^{14} {\rm M}_{\odot} \le M_{\rm CL} \le 
6.5 \times 10^{14} {\rm M}_{\odot}$ and the number fraction
of ICGCs does not depend on $M_{\rm CL}$ for the above
cluster mass range.

(2) The projected radial density profiles 
(${\Sigma}_{\rm GC}$) of ICGCs
in clusters with different $M_{\rm CL}$
can be diverse,
though ICGCs have inhomogeneous, asymmetric,
and somewhat elongated distributions in most models.
If ${\Sigma}_{\rm GC} (R) \propto R^{\alpha}$,
$\alpha$ ranges from $\approx -1.5$  to $\approx -2.5$ for GCs in clusters with
the above mass range.

(3) Although total number of GCs within the central 0.05 Mpc ($N_{\rm GC,0.05}$)
and 0.2 Mpc ($N_{\rm GC,0.2}$) are diverse in different clusters,
they can depend
weakly on $M_{\rm CL}$ in such a way that both $N_{\rm GC,0.05}$
and $N_{\rm GC,0.2}$ are likely to be
larger for clusters with larger $M_{\rm CL}$.

(4) Total number of GCs per cluster masses (specific frequency of GCs for
clusters of galaxies) are more likely to be larger in more massive
clusters,  mainly because a larger number of earlier virialized objects 
can be located in more massive clusters.

(5) Radial density distributions of all GCs including ICGCs and galactic
GCs have steeper profiles than those of the background dark matter 
halos in the central regions of clusters ($R<200$kpc)
with different  $M_{\rm CL}$.

(6) Spatial distributions of old GCs in clusters can depend on
the truncation epoch of GC formation ($z_{\rm trun}$) such that
they can be  steeper and more compact in the models 
with higher $z_{\rm trun}$.

(7) The mean metallicity of ICGCs in a cluster can be smaller than that
of GCs within the cluster member galaxy-scale halos by  $\sim 0.3$
in [Fe/H].
Metallicity distribution functions (MDFs)  of ICGCs show 
peak values around [Fe/H] $\sim -1.6$ 
and  do not have 
remarkable  bimodality.

\end{abstract}

\begin{keywords}
globular clusters: general -- galaxies: star clusters ---
galaxies: elliptical and lenticular ---
galaxies:evolution --
galaxies: interactions
\end{keywords}

\section{Introduction}

Structural, kinematical, and chemical
properties of globular cluster systems (GCSs)
have long provided valuable clues to the better understanding 
of how galaxies form and evolve (e.g.,  Harris 1991; West et al. 2004).
In particular,  
the specific frequencies ($S_{\rm N}$), colour bimodality, and
structural properties of GCSs have been discussed extensively
in many different contexts of galaxy formation
(e.g., Searle \& Zinn 1978; 
Forbes et al. 1997; Ashman \& Zepf 1998;
Brodie et al. 1998;  C\^ote et al. 2001; Beasley et al. 2002; 
Bekki et al. 2002).
The physical properties of globular clusters (GCs)
however have not been so far considered as fossil records that contain
valuable information on {\it how groups and clusters of galaxies were 
formed and evolved}. Furthermore,
GC properties characteristics of giants Es located in the central
regions of  clusters 
(e.g., very high $S_{\rm N}$) 
have not been extensively discussed in the context of 
cluster formation processes 
via hierarchical merging of smaller groups and clusters of galaxies. 
 
\begin{table}
\centering
\begin{minipage}{75mm}
\caption{Meaning of acronym}
\begin{tabular}{|l||l|} 
INGC &  INtergalactic GCs \\
ICGC &  IntraCluster GCs \\
IGGC &  IntraGroup GCs \\
GGC &  Galactic GCs \\
\end{tabular}
\end{minipage}
\end{table}

\begin{table*}
\centering
\begin{minipage}{185mm}
\caption{Brief summary of the results for each model}
\begin{tabular}{ccccccccc}
(1)&(2)&(3)&(4)&(5)&(6)&(7)&(8)&(9)\\
{Model no.  
\footnote{The results of the cluster-scale halos with the total
masses ($M_{\rm CL}$) larger than $10^{14} {\rm M}_{\odot}$ are described.}} &
$M_{\rm CL}$ ($\times 10^{14} {\rm M}_{\odot}$)  & 
{$N_{\rm GC,CL}$ 
\footnote{Total number of GCs in a cluster-scale halo.}} &
{$S_{\rm N,CL}$%
\footnote{Specific frequency of cluster GCs. 
This is defined as total GC number per mass for a cluster of
galaxies and given in  units of 
$ {\rm number} / 10^{10} {\rm M}_{\odot}$.}} &
$R_{10}$   (Mpc) & 
$R_{50}$   (Mpc) & 
$f_{\rm ICGC}$  & 
{$N_{\rm GC,0.05}$ %
\footnote{Total number of GCs within 0.05 Mpc.}} &
{$N_{\rm GC,0.2}$ %
\footnote{Total number of GCs within 0.2 Mpc.}} \\
CL1 &  6.5 & 6868 &  0.11 &  0.026 &
0.76 &  0.29 & 1098 & 2084 \\
CL2  & 3.1 & 3449 &  0.11 &  0.014 &  0.21 &   
0.31 &  962 &  1699 \\
CL3  & 2.9 & 3248  & 0.11 &  0.026 &  0.51 &  
  0.40 &  479 & 947 \\
CL4  & 2.3 & 3788 & 0.17 & 0.012 & 0.12 & 
  0.18 & 1348 & 2262 \\
CL5  & 2.0 & 2251  & 0.11 &  0.013 &  0.50  & 
  0.27 & 597 & 944 \\
CL6  & 1.8 & 2214 &  0.13 & 0.011  & 0.33 & 
  0.24 & 766 & 1052 \\
CL7  & 1.7 & 2197 &  0.13 &  0.068 &  0.32 & 
  0.37 & 145 &  762\\
CL8  & 1.5 & 1905 &  0.12 & 0.030 &  0.23 & 
  0.27 & 292 & 883 \\
CL9  & 1.4 & 1685 &  0.12 &  0.013 &  0.24  & 
  0.25 & 470 & 801 \\
CL10 &  1.3 & 1649 &  0.13 &  0.028  &  0.31 &  
  0.29 & 311 & 717 \\
CL11 &  1.2 & 1347 &  0.11 &  0.028 &  0.23 &  
  0.29 & 224 & 620 \\
CL12 &  1.1 & 795 &  0.07 &  0.035 &  0.37 &   
 0.35 & 104 & 235 \\
CL13  &  1.0 & 1580  & 0.15 &  0.029 &  0.33 & 
  0.31 & 248 & 640 \\
CL14 &  1.0 &  1042 &  0.10 &  0.028 &  0.34 & 
  0.28 & 186 & 444 \\
\end{tabular}
\end{minipage}
\end{table*}

Observational studies of GCs 
in clusters of galaxies have suggested that
there can be a population of GCs that are 
bounded by cluster gravitational potentials 
rather than those of cluster member
galaxies (e.g., White 1987; Bassino et al. 2002, 2003; 
Jord\'an et al. 2003).
Mar\'in-Franch \& Aparicio (2003) investigated
whether these intracluster GCs (``ICGCs'')  exist in the Coma cluster
using surface brightness fluctuations,
and concluded that ICGCs are highly unlikely to exist in the Coma.
West et al. (1995) suggested that the observed large variations 
in GC number (or $S_{\rm N}$)
between central giant Es in different clusters are due
to the existence of a population of GCs that are not bounded 
to individual galaxies but move freely throughout the core of
clusters galaxies.
If very high $S_{\rm N}$ in central Es in some clusters (e.g., Abell 2052)
are due to  larger number of ICGCs in the clusters,
it is an essentially important question why only some clusters of galaxies
contain larger number of ICGCs.

A new type of sub-luminous and extremely compact ``dwarf galaxy'' has
been recently discovered in an `all-object'  spectroscopic survey
centred on the Fornax Cluster (Drinkwater et al. 2000; 2003).
These ``dwarf galaxies'', which are members of the Fornax Cluster,
have intrinsic sizes of $\sim$ 100 pc and absolute
$B$ band magnitude  ranging from $-13$ to $-11$ mag
and are thus called ``ultra-compact dwarf'' (UCD) galaxies.
Structural and kinematical studies of this new population of
very bright ``star clusters''
have also suggested  that these clusters can be
also freely floating intracluster objects 
(Mieske et al. 2004; Drinkwater et al. 2005; Jones et al. 2005).
The observed compact spatial distribution and smaller velocity
dispersion of this possible intracluster population in the
Fornax cluster has not been clearly explained in a self-consistent
manner by previous theoretical studies (e.g., Jones et al. 2005).

Physical properties of intracluster stellar objects 
such as ICGCs and PNe 
are considered to be sensitive to 
dark matter properties and
cluster-related physical processes (e.g., tidal stripping
of GCs and hierarchical growth of clusters)
and thus provide some fossil information on formation of
galaxies and clusters of galaxies 
(e.g., Arnaboldi 2004 for a recent review).
However 
there has been little theoretical and numerical works  carried out 
as to how ICGCs are formed in clusters environments 
(e.g., Forte et al. 1982; Muzzio 1987; Bekki et al. 2003). 
These previous models showed that 
tidal stripping of GCs from cluster member galaxies
though galaxy-galaxy and galaxy-cluster interaction
is a mechanism for ICGC formation.
These previous works however used {\it fixed} gravitational potentials of 
already virialized clusters 
and accordingly could not discuss {\it how ICGCs in  a cluster  are formed as
the cluster grows through hierarchical merging of smaller groups
and clusters.}

These previous observational and theoretical studies have
raised many questions, the most significant being: 
(i) How are ICGC  formed 
under the currently
favored cold dark matter (CDM) theory of galaxy formation ?
(ii) What physical properties can  ICGCs  have if their formation
is closely associated with hierarchical formation of clusters ?
(iii) Are there any correlations between physical properties
of ICGCs and global properties (e.g., masses and X-ray temperature)
of their host clusters of galaxies ?
(iv) Is the origin of very high $S_{\rm N}$ of central giants Es
in some clusters closely associated with the formation processes
of clusters of galaxies (e.g., galaxy accretion from fields) ?
and (v) Are there any differences in radial distributions
and kinematics between UCDs and ICGCs ?

In order to answer these questions in a fully self-consistent manner,
we need to model both (1) dynamical evolution of GCs  (e.g., tidal stripping
and accretion of GCs) during hierarchical growth of clusters of galaxies
and (2) formation of GCs at low and high redshifts in clusters environments.
Although recent high-resolution cosmological N-body simulations with
total particle number of $ \sim 10^8$ 
(e.g., Yahagi \& Bekki 2005; YB) can allow us to address the above point (1),
our poor understanding 
of the above point (2) on
physical conditions required for GC formation in galaxies 
(e.g., Ashman \& Zepf 1998) would make it difficult for us to
have robust conclusions for the above five problems.
However,  considering recent significant development in
observational studies on GCs in nearby clusters
(e.g., Dirsch et al. 2003; Jord\'an et al. 2003; Richtler et al. 2004),
it is still doubtlessly worthwhile to provide  
some theoretical predictions that can be tested against these latest
and future observational results.

The purpose of this paper  is thus to investigate extensively 
physical properties of GCs in clusters of galaxies
based on high-resolution cosmological simulations that can
follow both hierarchical growth of clusters through merging of
smaller subhalos 
and dynamical evolution of old GCs.
We particularly investigate global ($\sim$ Mpc scale) 
 density distributions and
kinematics both for ICGCs and for GCs that are within galaxy-scale
halos in clusters (referred to as ``GGCs'' for convenience).
Because of some uncertainty in the modeling of the formation
processes of young, metal-rich
GCs (e.g., Bekki et al. 2002),
we investigate physical properties of {\it cluster GCs
that were formed at high redshift $z>6$}.
Accordingly, the present study is regarded as  only the first step
toward better understanding the nature of GCs in clusters of 
galaxies: The results of our simulations may well be compared with
observations on ``blue GCs'' in  a more reasonable way.

The plan of the paper is as follows: In the next section,
we describe our  numerical models for dynamical evolution
of GCs in forming clusters of galaxies. 
In \S 3, we
present the numerical results
mainly on radial distributions  and kinematics of GCs in clusters
with different masses. 
In \S 4, we discuss (1) the origin of very high $S_{\rm N}$
of giant Es in the central regions of some clusters,
(2) the origin of UCDs, and (3) the importance of the epoch
of the truncation of GC formation by reionization
in spatial distributions of cluster GCs. 
These points were not discussed at all in YB.
We summarize our  conclusions in \S 5.

Throughout this paper,
we use ICGCs rather than ICG (West et al. 1995) in order to distinguish
between intragroup GCs (IGGCs) and intracluster GCs (ICGCs).
GCs within any galaxy-scale halos in a cluster-scale halo
are referred to as GGCs so that they can be distinguished from
ICGCs 
(Note that these GGCs are {\it not} the Galactic
GCs).
GCs that were formed
within subhalos at $z>6$ yet are not within any virialized halos
at $z=0$ can be regarded as ``intergalactic''
(van den Bergh 1958)
or ``intercluster'' (``intergroup'') GCs (YB).
Intergalactic, intergroup, and intercluster GCs
are simply referred to as INGCs,
because these three GC populations would be difficult
to be distinguished with one another in observations.
For clarity and convenience, 
the meanings of these acronym are given in Table 1.

\section{The model}

We first identify hypothetical ``GC particles''  at high redshifts ($z>6$)
in the high-resolution,
collisionless cosmological N-body simulation
and then follow their dynamical evolution till $z=0$.
We mainly investigate structural and kinematical properties of
the simulated ``GC particles'' that are located in cluster-scale 
halos with $M_{\rm CL} \ge 10^{14} {\rm M}_{\odot}$ at $z=0$.

\subsection{Identification of ``GC'' particles}

We simulate the large scale structure of GCs  
in a $\Lambda$CDM Universe with ${\Omega} =0.3$, 
$\Lambda=0.7$, $H_{0}=70$ km $\rm s^{-1}$ ${\rm Mpc}^{-1}$,
and ${\sigma}_{8}=0.9$ 
by using the Adaptive Mesh Refinement $N-$body code developed
by Yahagi (2005) and Yahagi et al. (2004), 
which is a vectorized and parallelized version
of the code described in Yahagi \& Yoshii (2001).
We use $512^3$ collisionless dark matter (DM) particles in a simulation
with the box size of $70h^{-1}$Mpc and the total mass 
of $4.08 \times 10^{16} {\rm M}_{\odot}$. 
We start simulations at $z=41$ and follow it till $z=0$
in order to investigate physical properties
of old GCs outside and inside virialized dark matter halos at $z=0$. 
We used the COSMICS (Cosmological Initial Conditions and
Microwave Anisotropy Codes), which is a package
of fortran programs for generating Gaussian random initial
conditions for nonlinear structure formation simulations
(Bertschinger 1995, 2001). 

\begin{figure*}
\psfig{file=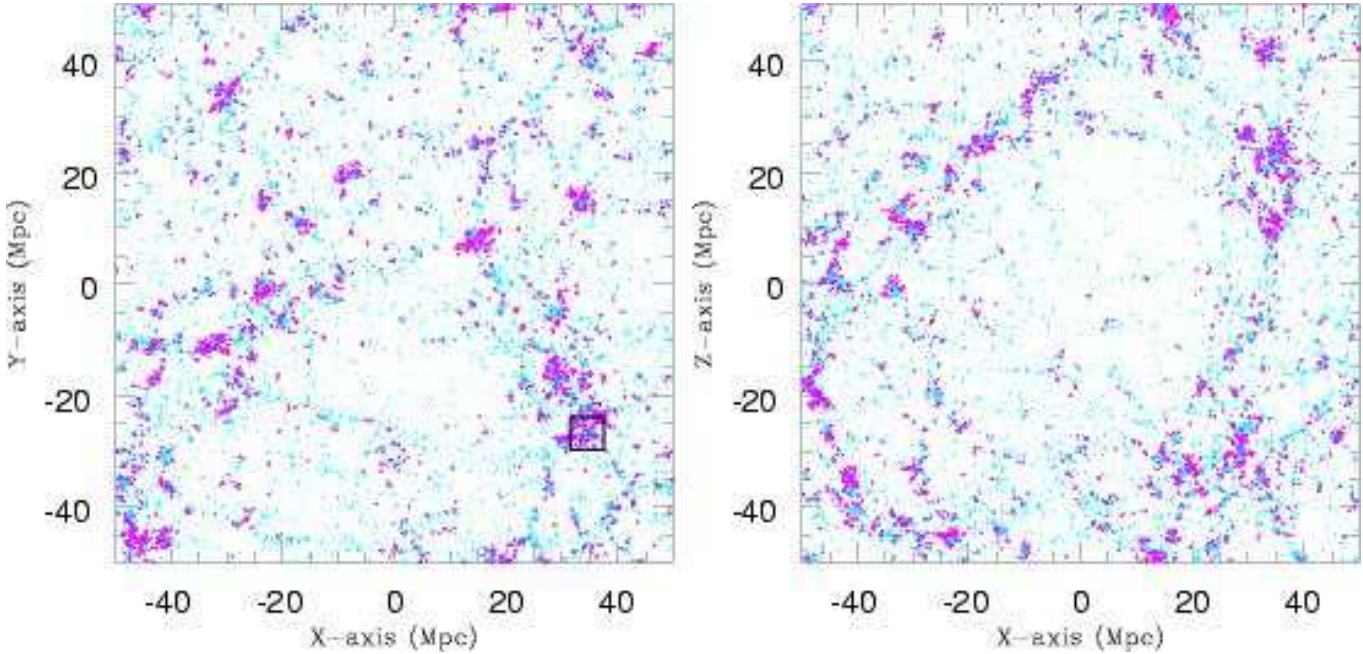,width=18.cm}
\caption{
The large scale structure of old GCs formed at $z>6$ 
projected onto the $x$-$y$ plane (left)
and the $x$-$z$ one (right).
GCs that are not within any virialized halos at $z=0$ (INGCs)
are plotted by bigger magenta dots whereas
those within halos are plotted by cyan dots.
A frame measures  100 Mpc for the two panels. 
The square region shown by a thick solid line  is the region
where GCs in the fiducial cluster model (CL1) are located. The detailed
distribution of GCs in the CL1 model is shown in Fig. 2. 
}
\label{Figure. 1}
\end{figure*}

The way of investigating GC properties is described as follows.
Firstly we select virialized dark matter subhalos at $z=z_{\rm trun}$ 
by using the friends-of-friends (FoF) algorithm (Davis et al. 1985)
with a fixed linking length of 0.2 times the mean DM particle separation.
For each individual virialized subhalo
with the half-mass radius of $R_{\rm h}$,
some fraction ($f_{\rm gc}$) of particles within $R_{\rm h}/3$ are labeled 
as ``GC'' particles. This procedure for defining GC particles
is based on the assumption that energy dissipation via radiative cooling
allows baryon to fall into the deepest potential well of dark-matter halos
and finally to be converted into GCs.
The value of $R_{\rm h}/3$ is chosen, because the size 
of the Galactic GC system  is similar to $R_{\rm h}/3$ of the dark matter
halo in the dynamical model of the Galaxy (Bekki et al. 2005).
We assume that old, metal-poor globular cluster formation is truncated
after $z=z_{\rm trun}$, because previous theoretical studies
demonstrated that such truncation of GC formation 
by some physical mechanisms (e.g., reionization) is necessary
for  explaining  the color bimodality of GCs,
very high specific frequency ($S_{\rm N}$)
in cluster Es,  and structural properties of the Galactic 
old stars and GCs (e.g., Beasley et al. 2002; Santos 2003;
Bekki 2005; Bekki \& Chiba 2005).

Secondly we follow the dynamical evolution of GC particles
formed before $z=z_{\rm trun}$ till $z=0$ and thereby
derive locations ($(x,y,z)$) 
and velocities ($(v_{\rm x},v_{\rm y},v_{\rm z})$)
of GCs at $z=0$.
We then identify virialized halos at $z=0$ with the FoF algorithm
and investigate whether each of GCs is within the halos.
If a GC is found to be within a halo, the mass of the host halo
($M_{\rm h}$)
and the distance of the GC from the center of the halo 
($R_{\rm gc}$) are investigated.
If a GC is not in any halos, it is regarded as an 
intergalactic GC (``INGC'')
and the distance ($R_{\rm nei}$) between the INGC and the nearest neighbor
halo and the mass of the halo ($M_{\rm h,nei}$) are
investigated.
If a GC is found to be within a cluster-size halo 
($M_{\rm h} > 10^{14} {\rm M}_{\odot}$), 
we investigate which galaxy-scale halo in the cluster-scale
halo contains the GC.
If we find the GC within the tidal radius of 
one of galaxy-scale halos 
in the cluster-scale halo, 
it is regarded as a galactic GC (GGC):
Otherwise it is regarded as an ICGC. 
The way to select galaxy-scale halos within  a cluster-scale halo
is given in the Appendix A.
Furthermore the details of
the method  to determine the tidal radius of the GCS in each
galaxy-scale halo in a cluster are  given and discussed 
in the Appendix B.

Thus, the present simulations enable us to investigate
physical properties only for old  GCS
owing to the adopted assumptions of collisionless simulations:
Physical properties of metal-rich
GCs lately formed during secondary dissipative galaxy  merger events
at lower redshifts (e.g., Ashman \& Zepf 1992) can not
be predicted by this study.
We present the results of the model with $z_{\rm trun}=6$,
10, and 15.
If $z_{\rm trun}$ is closely associated with the completion
of cosmic reionization, $z_{\rm trun}$ may well range
from 6 (Fan et al. 2003)
to 20 (Kogut et al. 2003).
Physical properties of hypothetical GC particles for
ICGCs in clusters with
$M_{\rm h} \ge  10^{14} {\rm M}_{\odot}$
are described for
the models with $f_{\rm gc}=0.2$ in which
the number ratio  of GC particles to all particles
is $1.5 \times 10^{-3}$ at $z=0$.

This value of 0.2 for $f_{\rm gc}$
was chosen such that the typical subhalo at z=6
($M_{\rm h} \sim   1.8 \times  10^{10} {\rm M}_{\odot}$) 
can have at least one ``GC particle'' within
$R_{\rm tr,gc}$ in the simulation 
with the mass resolution of  $\sim   3 \times  10^{8} {\rm M}_{\odot}$.
The present results do not
depend on $f_{\rm gc}$ at all as long as $f_{\rm gc}\ge 0.1$.
The dependences of the present results on $f_{\rm gc}\ge 0.1$
are discussed  in the Appendix C.
The minimum particle number ($N_{\rm min}$)
that is required for halo identification is
set to be 10.

We assume that the initial radial ($r$) profiles of GCSs ($\rho (r)$) in
subhalos at $z=6$ are the same as those of the simulated dark matter
halos that can have the universal ``NFW'' profiles
(Navarro, Frenk, \& White 1996) with
$\rho (r) \propto r^{-3}$ in their outer parts.
The mean mass of subhalos at z=6 in the present simulations is roughly
$1.8 \times  10^{10} {\rm M}_{\odot}$,
which is similar to the total mass
of bright dwarf galaxies.
Minniti et al. (1996) found that the projected ($R$) density
profiles of GCSs in dwarfs is approximated as $\rho (R) \propto R^{-2}$, which
is translated roughly as  $\rho (r) \propto r^{-3}$
by using a canonical conversion formula from
$\rho (R)$ into $\rho (r)$ (Binney \& Tremaine 1987).
Therefore, the above assumption on $\rho (r)$ can be regarded as reasonable.
Thus we mainly show the fiducial model with $\rho (r)$ similar
to the NFW profiles and $R_{\rm tr,gc}=R_{\rm h}/3$.

Although we base our GC models
on {\it observational results of GCSs at z=0},
we  can not confirm whether
the above
$\rho (r)$ and  $R_{\rm tr,gc}$
of the fiducial model are
really the most probable  (and the best)
for  GCSs {\it for low-mass subhalos at z=6 owing to the  lack of
observational studies of GCSs at high redshifts}.
As our previous paper (YB) showed,
the numerical results can depend on
initial $\rho (r)$ and  $R_{\rm tr,gc}$ of subhalos
at $z=6$. For example, the number fraction of INGCs
can be a factor of $\sim 2$ larger in the models with
$R_{\rm tr,gc}=R_{\rm h}/3$ than in those with 
$R_{\rm tr,gc}=R_{\rm h}/6$ both for the NFW and the power-law GC density
profiles.  The number fraction of INGCs or ICGCs (IGGCs) can be larger 
for models with less centrally concentrated initial GC distributions
in the present study. These suggest that
the observed properties of INGCs and ICGCs (IGGCs) can give some
constraints on initial GC distributions in subhalos at high $z$,
if they are compared with the corresponding simulations as
those done in the present study.

\subsection{``GN'' particles}

We consider that a GC particle in the very center of a galaxy-scale
halo at $z=z_{\rm trun}$  is a ``galactic nucleus'' (GN)  particle:
The central particle  
within $R_{\rm h}/3$ of each individual virialized subhalo 
at $z=6$ is  labeled as GN and other particles within
$R_{\rm h}/3$ are labeled as GCs.
It should be stressed here that 
although we follow dynamical evolution of GCs and GNs
till $z=0$,
{\it both GCs and GNs at $z=6$ are regarded as GCs at $z=0$}.
We do not discriminate GC and GN particles in most investigation,
because  (1) results are not so remarkably different between
GCs and GNs 
and (2) it would be  very hard to observationally discriminate between
GCs and (stripped)  GNs owing to the observed similarity in physical properties
of GCs and GNs (e.g., Walcher et al. 2005). 
However, in order to discuss the origin of dE,Ns and UCDs in \S 4, 
we try to 
investigate {\it separately}  some physical properties
of GCs at $z=0$ originating from GNs at $z=6$ and 
those from GCs at $z=6$.
Thus only some key results on GN particles  
(e.g., radial distributions of GNs at $z=6$)
are described in \S 3.3.

Number ratio of GN to GC particles ($f_{\rm gn}$) is typically 0.1
in the model with  $z_{\rm trun}=6$ for the cluster-scale halos
with $10^{14} {\rm M}_{\odot} \le M_{\rm CL}$.
However, the value of this $f_{\rm gn}$ should be carefully
compared with the corresponding observations, because one GC particle
(with the mass equal to $3.0 \times 10^8 {\rm M}_{\odot}$)
does not mean one GC in the present study.
Total number of GCs in the simulation is 207081 (for $f_{\rm gc}=0.2$), 
which means that
the GC number per mass (${\epsilon}_{\rm sim}$) is $5.1 \times 10^{-12}$
${\rm number}/{\rm M}_{\odot}$.
McLaughlin (1999) showed that total number of initial GCs
in a galaxy
can decrease by a factor of 25 within the Hubble time
owing to GC destruction by the combination effect of
galactic tidal fields and internal GC evolution
(e.g., mass loss from massive and evolved stars). 
This means that the initial GC number per
mass (${\epsilon}_{\rm obs}$)
is about $1.8 \times 10^{-9}$
for the Galaxy with 
the  mass of $\approx$ $2.0$ $\times$ $10^{12}$ $M_{\odot}$
(Wilkinson \& Evans 1999) and the observed GC number of 140
(van den Bergh 2000).
Therefore it is quite reasonable to consider that 
$f_{\rm gn} \times {\epsilon}_{\rm sim}/{\epsilon}_{\rm obs}$
(typically $ {\epsilon}_{\rm sim}/{\epsilon}_{\rm obs}\sim 0.003$)
rather than simple $f_{\rm gn}$
for each cluster-scale halo
is the value that should be compared  with  observation. 

\begin{figure}
\psfig{file=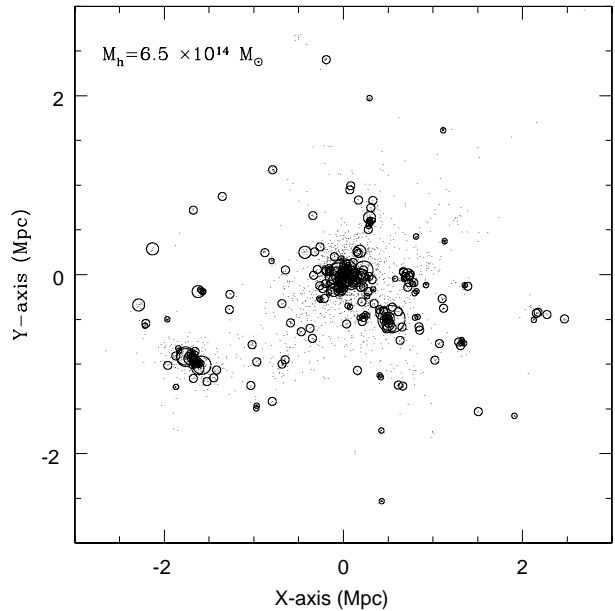,width=8.cm}
\caption{
Distributions of GCs projected onto the $x$-$y$ plane
at $z=0$ for the fiducial model CL1 
with the total mass of $6.5 \times 10^{14} {\rm M}_{\odot}$.
GCs within circles represent those within tidal radii of
galaxy-scale halos (GGCs) and the radii of the circles
represent the tidal radii.
GCs that are not within any circles are regarded as 
intracluster GCs (ICGCs).
}
\label{Figure. 2}
\end{figure}

\begin{figure}
\psfig{file=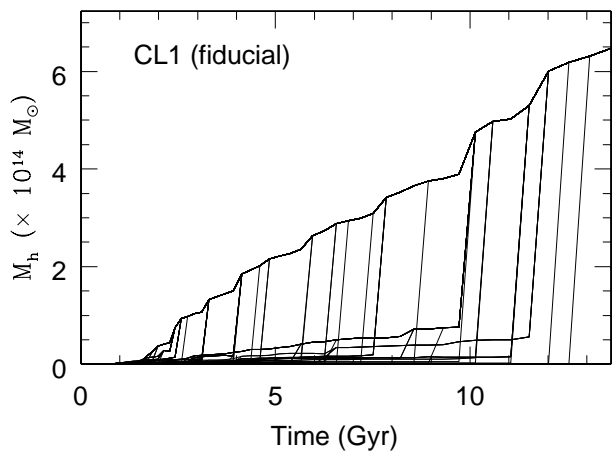,width=8.cm}
\caption{
Time evolution of total masses of halos that
are virialized at $z \ge 6$,
have $M_{h} \ge 10^{11} {\rm M}_{\odot}$ at $z=6$,
and belong to the cluster of galaxies at $z=0$ 
in the fiducial
cluster model CL1.
Each line represents the mass of a halo from 
$z=6$ to $z=0$.
This figure therefore shows the mass growth history of
the cluster via hierarchical clustering.
}
\label{Figure. 3}
\end{figure}

\begin{figure*}
\psfig{file=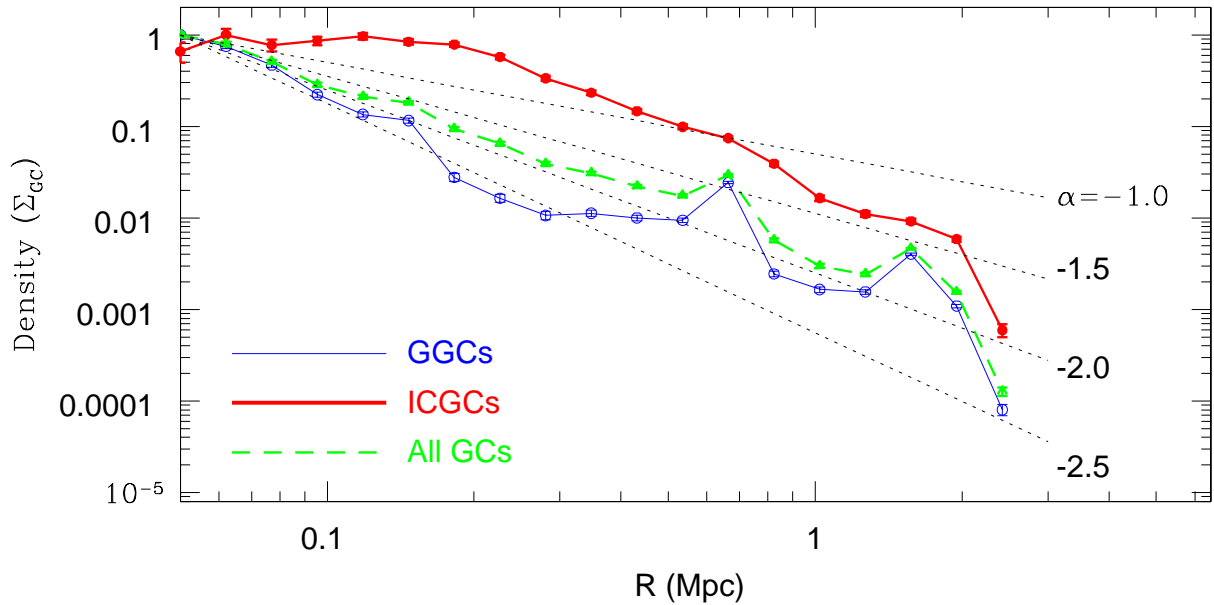,width=16.cm}
\caption{
Projected radial density profiles of GCs (${\Sigma}_{\rm GC}$)
for all GGCs (blue thin
solid), ICGCs (red thick solid), and all GCs (green thick dashed) 
in the cluster CL1. 
For clarity, the density distributions are normalized to
their maximum values.
Thin dotted lines represent power-law slopes ($\alpha$)
of $\alpha$ = $-2.5$, $-2.0$, $-1.5$, and $-1.0$.
}
\label{Figure. 4}
\end{figure*}

\begin{figure}
\psfig{file=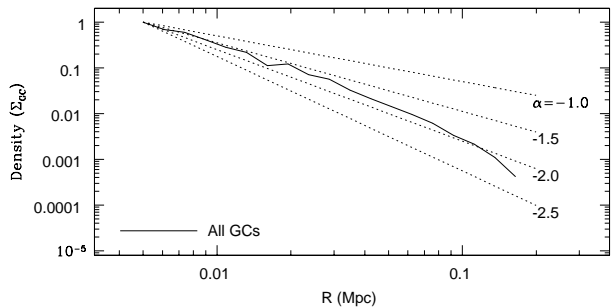,width=8.cm}
\caption{
The same as Fig. 4 but for all GCs within the central 200 kpc
of the cluster CL1.
}
\label{Figure. 5}
\end{figure}

\begin{figure}
\psfig{file=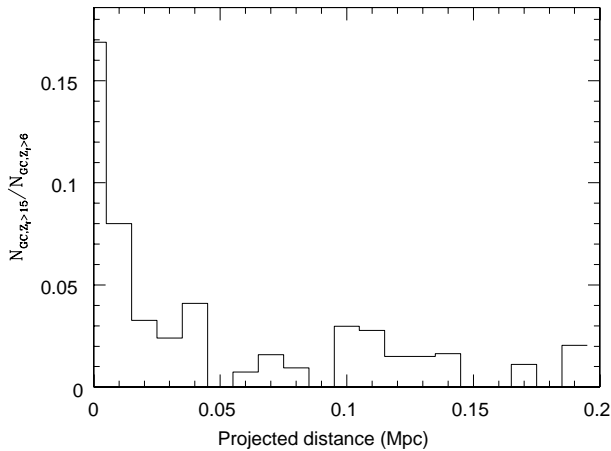,width=8.cm}
\caption{
The radial gradient of the GC number ratio of $N_{\rm GC, Z_{\rm f}>15}$
to $N_{\rm GC, Z_{\rm f}>6}$ in the CL1, where
 $N_{\rm GC, Z_{\rm f}>15}$ and $N_{\rm GC, Z_{\rm f}>6}$
represent GCs formed before $z=15$ and $z=6$, respectively.
This figure accordingly shows the age gradient of GCs in the central
region of the cluster.  
}
\label{Figure. 6}
\end{figure}

\begin{figure}
\psfig{file=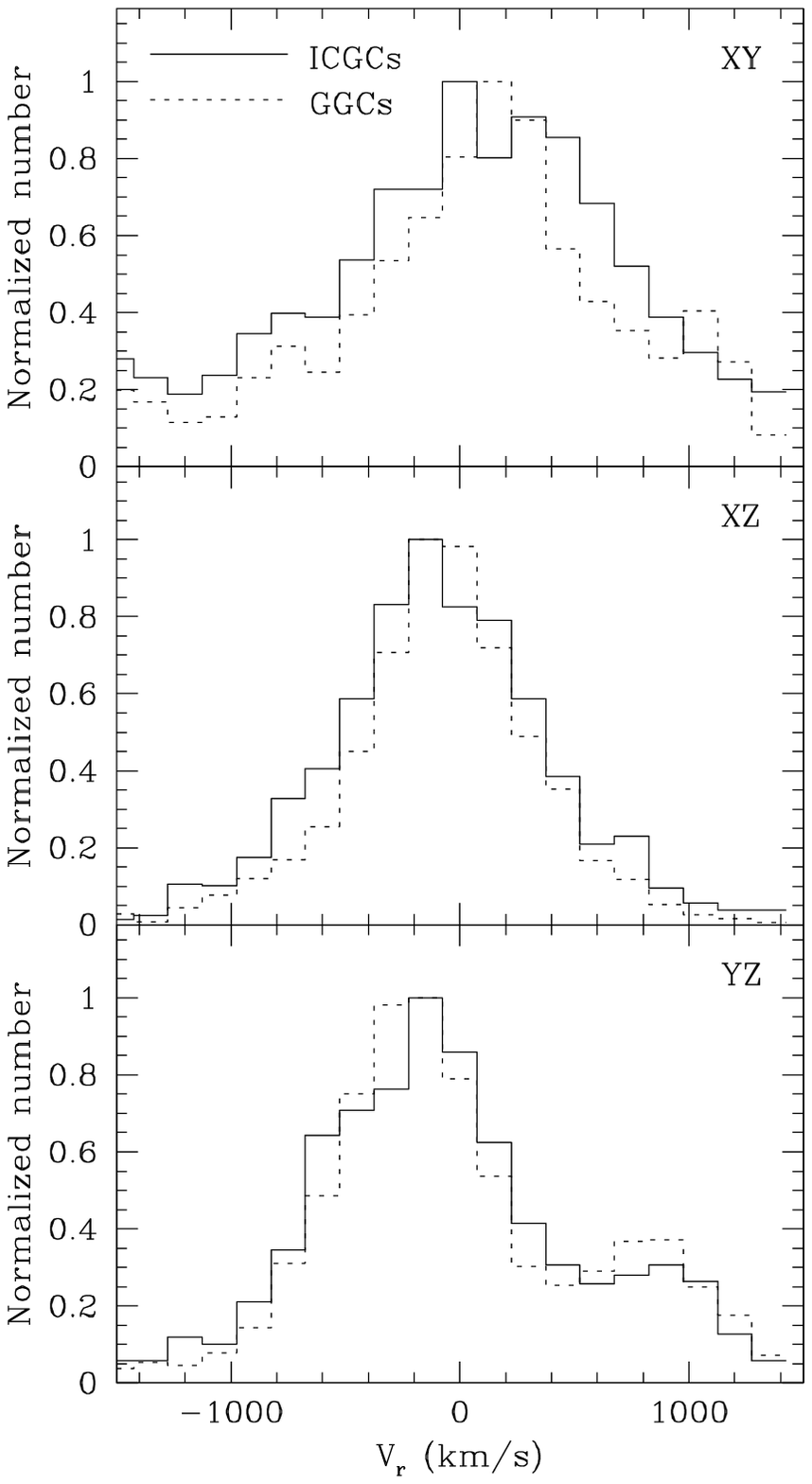,width=8.cm}
\caption{
Distribution of radial velocities of ICGCs (solid)
and GGCs (dotted)  projected onto 
the $x$-$y$ plane (top), the $x$-$z$ one (middle), and
the $y$-$z$ one (bottom) for the CL1. 
For clarity and comparison,
the number of GCs normalized to maximum values for each GC
population is shown for  each projection.
}
\label{Figure. 7}
\end{figure}

\subsection{Main points of analysis}

We select 14 virialized cluster-scale halos 
with the total virialized masses ($M_{\rm h}$) larger
than $10^{14} {\rm M}_{\odot}$ at  $z=0$ from the abovementioned
cosmological N-body simulation in which
27 cluster-scale halos can be formed
within the Hubble time.
Although we have already described some results 
of the most massive cluster model in YB,
this paper first describes dependences
of the physical properties of ICGCs 
on  $M_{\rm h}$ for different 14 cluster models.
Therefore, we can discuss whether the results in YB can
be also true for models with different cluster masses
in the present study.

We first  describe some generic results and then those
that  depend on  $M_{\rm h}$ and model parameters.
We first describe 
physical properties of hypothetical GC particles for
GGCs and ICGCs in the ``fiducial model'' with 
$M_{\rm h} = 6.5 \times 10^{14} {\rm M}_{\odot}$ (\S 3.1),
which show some generic results on radial density profiles and kinematics
of GGCs and ICGCs.
Then we show  the  dependences of the results on
$M_{\rm h}$ and $z_{\rm trun}$ (\S 3.2).
In particular, we focus on how the total number and 
radial density profiles of GCs in
the central regions of clusters can depend on $M_{\rm h}$,
because recent wide-field photometric study of GCs in
clusters (e.g., Dirsch et al. 2003) have begun to reveal
global structure and kinematics of GCs in the central regions
of clusters.

Table 2 summarises briefly the results for each  model:
Model number (column 1), 
total masses of cluster-scale halos $M_{\rm CL}$ (2),
total number of GCs in the clusters  $N_{\rm GC, CL}$ (3),
specific frequency of GCs for clusters  $S_{\rm N,CL}$ (4),
radius within which 10 \% of GCs are included $R_{10}$ (5), 
radius within which 50 \% of GCs are included $R_{50}$ (6), 
number fraction of ICGCs $f_{\rm  ICGC}$ (7),
total number of GCs within the central 0.05 Mpc $N_{\rm GC, 0.05}$ (8),
and total number of GCs within the central 0.2 Mpc $N_{\rm GC, 0.2}$ (9).
$S_{\rm N,CL}$ is defined as the total number of GCs per a cluster mass
and given in units of ${\rm number}/10^{10} {\rm M}_{\odot}$.
All of these results are for the model with $f_{\rm gc}=0.2$ and 
the derived parameter dependences on $M_{\rm CL}$ and $z_{\rm trun}$
does not depend on $f_{\rm gc}=0.2$.

\section{Results}

\subsection{The fiducial cluster model}

Fig. 1 shows the large scale ($\sim 100$Mpc) structure of
old GCs formed at $z>6$ 
for the model with $z_{\rm trun} = 6$ at $z=0$. 
The simulated GC large scale distribution 
shows strong clustering and 
appears to be quite similar to dark matter distributions simulated in previous
numerical works (e.g., Yahagi et al. 2002).
About 28 \% of all GCs are located within cluster-scale halos
with $M_{\rm h}  \ge  10^{14} {\rm M}_{\odot}$
About 99\% of all GCs formed before $z=6$ are located in galaxy-scale
or cluster-scale (or group-scale) halos at $z=0$, which means 
that only 1\% of the GCs can  be outside  any virialzied halos
and thus can be identified as intergalactic or intercluster
GCs (``INGCs''). The predicted physical properties of these INGCs
are described in detail in our previous paper  (YB).  

Fig. 2 shows the spatial distributions of GCs inside and outside galaxy-scale
halos in the fiducial cluster model
CL1 with $M_{\rm h}$ (or $M_{\rm CL}$) = $6.5 \times 10^{14} {\rm M}_{\odot}$.
About 29 \% of GCs in the cluster CL1 are not within any galaxy-scale
halos in the cluster so that they can be identified as ICGCs. These ICGCs
were GCs and GNs  of smaller galaxy-scale halos before $z=6$ and
tidally stripped from the halos to become ICGCs during the growth of the 
cluster via hierarchical merging of the halos.
The number fraction of GCs at $z=0$ that were previously
nuclei at $z > 6$ (i.e., GNs) is about 0.2, which implies that a significant
fraction of ICGCs can show physical properties atypical of normal
GCs (e.g., bimodal/multiple metallicity distributions 
seen in the most massive Galactic GC, $\omega$ Cen).
The ICGCs that were initially nuclei of relatively massive halos
($M_{\rm h} > 10^{10} {\rm M}_{\odot}$ corresponding to
host halos of bright dwarf galaxies with $M_{\rm B}<-16$ mag
for $M/L \approx 10$)
at $z > 6$  can be the possible
candidates of UCDs, because these halos (dwarfs) are observed to
contain bright nuclei (i.e., progenitors of UCDs). 

Fig.2 also shows that the distribution of ICGCs is inhomogeneous, asymmetric,
and somewhat elongated, in particular, in the outer part of the cluster,
where a substructure (corresponding to an infalling group of galaxies)
can be seen (i.e., around $(X,Y) = (-1.7, -1.0)$ Mpc)).
The asymmetric structure of ICGC distribution seen in the simulation
can be seen in the observation by Bassino et al. (2006) for
the central region of the Fornax  cluster.
The simulated elongated structure is similar to  
the observed distributions of intracluster stellar light
discovered by Zibetti et al. for clusters of
galaxies (2005) at $z\sim 0.25$. 
The simulated inhomogeneous and elongated ICGC distributions
can be seen in most models of the present study.

Fig. 3 clearly shows that the cluster grows slowly over the Hubble
time via {\it minor} merging or accretion of group-scale or galaxy-scale halos
rather than via  violent {\it major} merging with the mass ratios
of two groups and clusters as large as $0.5-1.0$. 
This suggests that tidal stripping of GCs 
initially within  smaller halos with GCs during
accretion of the halos  is one of the main mechanism  of
ICGC formation: Violent major merging 
that  can significantly
change spatial distributions  of GCs is a rare event in
this rich cluster.
The cluster finally can contain  about 3.4 \% of all GCs 
formed at $z>6$ in the simulation box.

Fig. 4 shows that there is a significant difference in the slope $\alpha$
of the projected GC distribution (i.e., ${\Sigma}_{\rm GC} \propto R^{\alpha}$)
between GGCs and INGCs: The  distribution
of ICGCs is significantly flatter than that of GGCs in the central
few hundreds kpc of the cluster. 
A least square fit to the simulation data shown in
Fig. 4 gives 
$\alpha \approx -2.4$ for GGCs and $\alpha \approx -0.7$ for ICGCs
in the central region of the cluster.
Since ICGCs are selected on the base of whether GCs are well outside
the tidal radii of GCSs of galaxy-scale halos  in the present simulation,
it would not be so difficult for observational studies
to select ICGCs from all GCs in clusters in the same way as the
present study does. It is accordingly doubtlessly worthwhile
to compare the results in Fig. 4 and those derived in future wide-field
imaging of cluster GCs (e.g., Richtler et al. 2004) 
to understand the origin of ICGCs. 

Fig.5  shows that ${\Sigma}_{\rm GC}$ of all GCs
in the central region of the cluster
CL1 is quite flat ($\alpha \approx -1.5$ for $R<$ 20kpc).
Since these central GCs can be observationally identified as GCs
within the central galaxy of a cluster (e.g., cDs and BCDs),
the result in Fig. 5 implies that the GCS of the central giant
galaxy in a cluster has  a flat ${\Sigma}_{\rm GC}$.
The derived $\alpha$ value of $-1.5$ is actually similar to those observed 
for very bright Es with $M_{\rm V} < -23$ mag,  most of which are
central giant Es in clusters (Ashman \& Zepf 1998).
The total number of GCs within the central 50 kpc
($N_{\rm GC, 0.05}$) and 200 kpc  ($N_{\rm GC, 0.2}$) 
is  1098 and 2084, respectively, 
which corresponds to 16 \% and 30\%, respetively,
of all GCs in the cluster CL1.
This centrally concentrated distribution of old GCs  is one
of generic results in the present study.

Fig. 6  shows  the radial distribution
of the number ratio of GCs formed at $z > 15$ ($N_{\rm GC,Z_{\rm f} > 15}$)
and those at $z > 6$ ($N_{\rm GC,Z_{\rm f} > 6}$)
in central 200 kpc of the cluster CL1.
This radial distribution, which can be regarded as an age gradient
of GCs,  has a negative gradient and thus suggests that 
the inner cluster GCs are more likely to be older than the outer ones.
This is probably because galaxy-scale halos that are formed and virialized
at higher redshifts and have higher mass densities 
can finally reside in the inner regions of clusters. 
However, the derived age gradient is not so large ($\sim 0.5$ Gyr for
the adopted cosmology) that it would be 
observationally difficult to prove it.

Fig. 7 shows the difference in the radial velocity ($V_{\rm r}$;
line-of-sight-velocity)  between GGCs and ICGCs for 
the three projections for the cluster CL1.
There are no remarkable differences in the $V_{\rm r}$ distributions
between the two GC populations, though ICGCs have only slightly larger
velocity dispersion than GGCs in the $x$-$y$ projection.
Both GC populations show asymmetric $V_{\rm r}$ distributions in
the $x$-$y$ and the $y$-$z$ projections owing to the presence
of infalling small groups of galaxies in this cluster CL1.
These results suggest that the  
differences in kinematics between GGCs and ICGCs
can not be so clearly observed.

\begin{figure}
\psfig{file=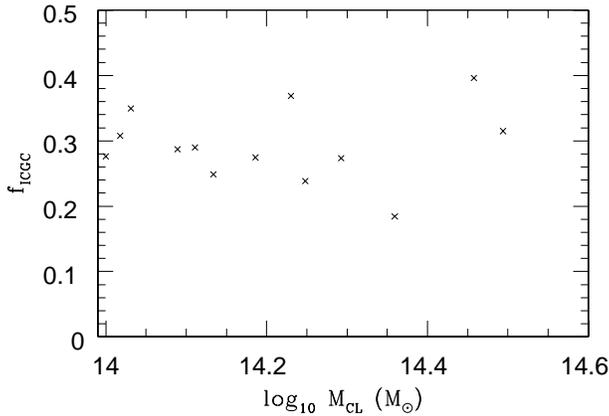,width=8.cm}
\caption{
Dependence of $f_{\rm ICGCs}$ 
on $M_{\rm CL}$ for the 14 clusters.
}
\label{Figure. 8}
\end{figure}

\begin{figure}
\psfig{file=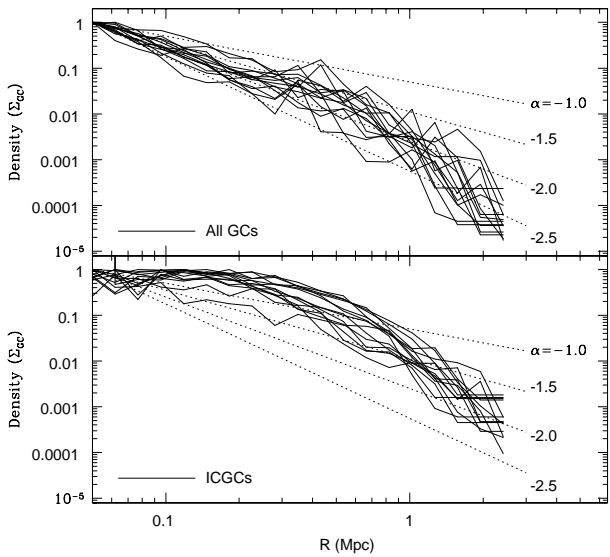,width=8.cm}
\caption{
The same as Fig. 4 but for all GCs 
(upper) and ICGCs (lower) in the 14 clusters.
}
\label{Figure. 9}
\end{figure}

\begin{figure}
\psfig{file=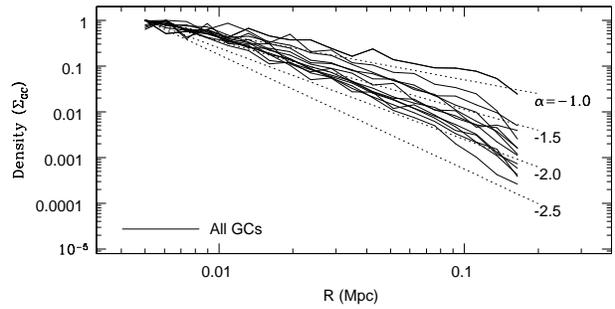,width=8.cm}
\caption{
The same as Fig. 5 but for all GCs 
in the 14 clusters.
}
\label{Figure. 10}
\end{figure}

\begin{figure}
\psfig{file=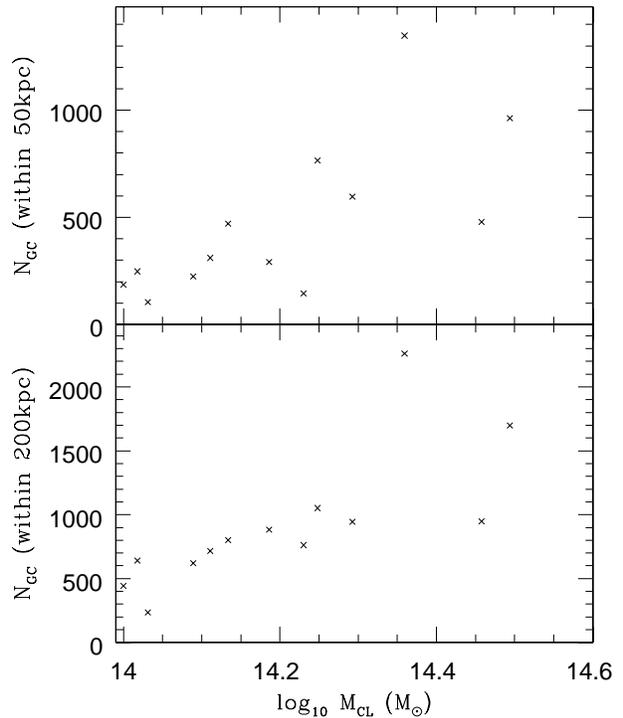,width=8.cm}
\caption{
Dependences of total number of GCs within the central 50 kpc (upper)
and 200kpc (lower) of clusters  on $M_{\rm CL}$.
Note that  a weak correlation between the GC number and $M_{\rm CL}$ 
can be seen in both panels.
}
\label{Figure. 11}
\end{figure}

\subsection{Parameter dependences}

Although the numerical results 
of {\it some} structural and  kinematical properties
of GCs  (e.g., flat ICGC distributions)
are similar to one another 
between the fiducial cluster model CL1 and  other  cluster models,
several physical properties of cluster GCs are quite diverse depending
on the cluster masses  ($M_{\rm CL}$) and the model parameters 
(e.g., $z_{\rm trun}$).
In Figs.  $8-15$, 
we illustrate  the derived dependences on  $M_{\rm CL}$ and $z_{\rm trun}$.

\subsubsection{$M_{\rm CL}$}

We find the following results:

(i) The number fraction of ICGCs ($f_{\rm ICGC}$) does not depend
on $M_{\rm CL}$ for the cluster mass range of 
1.0 $\le$ $M_{\rm CL}/10^{14} {\rm M}_{\odot}$ $\le$ 6.5 (Fig. 8).
The derived $f_{\rm ICGC}$ is diverse ranging from 0.18 to 0.40
and this diversity is due to the differences of merging histories
of clusters.

(ii) The projected radial density profiles (${\Sigma}_{\rm GC} (R)$)
for all GCs can be diverse with $ -1.5 < \alpha < -2.5$ in
the models with different $M_{\rm CL}$.
These differences are due to the differences in merging 
histories between clusters with different $M_{\rm CL}$ (Fig. 9).
Irrespective of $M_{\rm CL}$, ICGCs show flat profiles in the central 
$\sim$ 0.2 Mpc of clusters.
If we fit the profiles to 
$ {\Sigma}_{\rm GC} \propto {(1+R_{\rm c}/R})^{\alpha}$,
where $R_{\rm c}$ corresponds to the core radius of a profile,
we derive $ 0.03 \le R_{\rm c} \le 0.39$ (Mpc) and
$-4.0 \le \alpha \le -1.4$.

(iii) The projected radial density profiles of GCs for the central 100 kpc 
is more likely to be significantly flatter than those for the entire cluster
regions ($R<2$ Mpc). The slopes $\alpha$
of the inner GC profiles can become as flat as $-1.0$ (Fig. 10),
which is similar to those observed for the central giant Es in
some clusters. The derived slopes are quite diverse ranging from
$\approx -2.0$ to $\approx -1.0$ for the simulated cluster mass range. 

(iv) Both $N_{\rm GC,0.05}$ and $N_{\rm GC,0.2}$ are more likely
to be larger for larger $M_{\rm CL}$ (Fig. 11). This result implies that
more massive cluster are more likely to have higher densities of old
GCs. Also this suggests that central giants Es in more massive clusters
can have higher $S_{\rm N}$ of old GCs than those in less massive
clusters owing to the higher central GC number densities.
However there  are no clear trends in the relationships between
$M_{\rm CL}$, $R_{10}$, and $R_{10}/R_{50}$ (Fig. 12).

(v) The ratio of $N_{\rm GC}$ to $M_{\rm h}$ (or $M_{\rm CL}$),
which is referred to as
``GC specific frequency for clusters (groups)'' 
and represented as $S_{\rm N, CL}$),
depends weakly on $M_{\rm h}$ in the sense that $S_{\rm N, CL}$
is higher for more massive group-scale or cluster-scale halos (Fig. 13).
This is  because more massive groups/clusters are formed from
a larger number of smaller halos that are formed at $z>6$ 
(i.e., $z > z_{\rm trun}$) and thus can
have GCs. Larger dispersion of  $S_{\rm N, CL}$ can be seen
for groups and clusters with smaller masses. 

\begin{figure}
\psfig{file=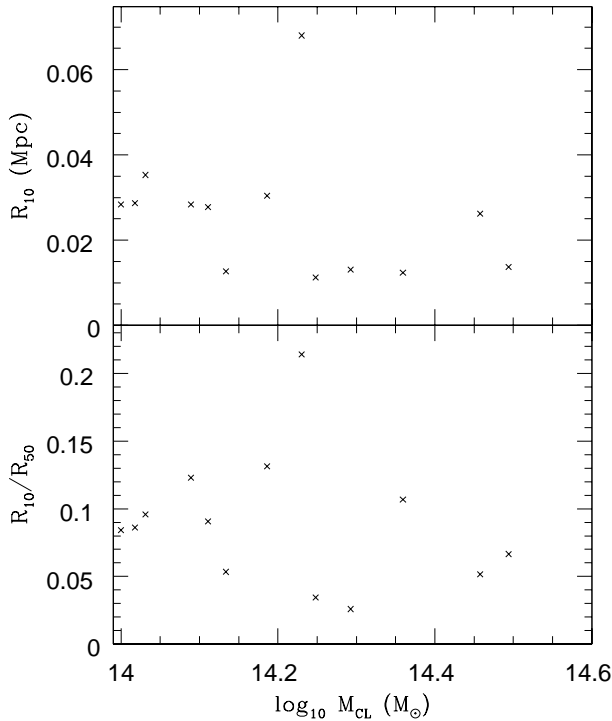,width=8.cm}
\caption{
Dependences of $R_{10}$ (upper)
and $R_{10}/R_{50}$ (lower) of GCs in  clusters  on $M_{\rm CL}$,
where $R_{10}$ and $R_{50}$ are the radii within which 10 \% and 50 \% of
all GCs are located, respectively.
Note that 
both $R_{10}$ and  $R_{10}/R_{50}$ have no remarkable 
correlations  with  $M_{\rm CL}$. 
}
\label{Figure. 12}
\end{figure}

\begin{figure}
\psfig{file=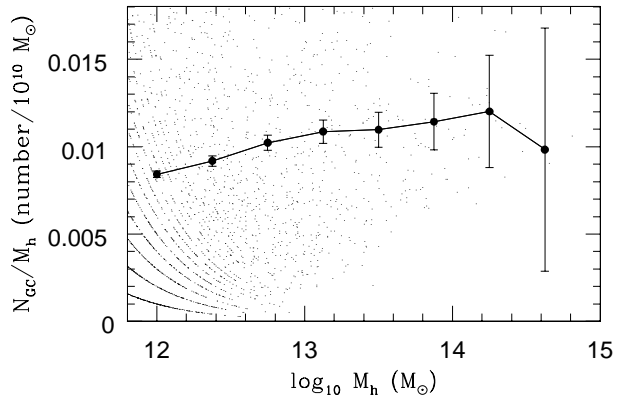,width=8.cm}
\caption{
The dependence of $N_{\rm GC}/M_{\rm h}$ on $M_{\rm h}$ 
for 
all group-scale ($M_{\rm h} \approx 10^{12}-10^{14} {\rm M}_{\odot}$)
and cluster-scale ($M_{\rm h} \approx 10^{14}-10^{15} {\rm M}_{\odot}$)
halos.
This ratio of  $N_{\rm GC}/M_{\rm h}$ on $M_{\rm h}$ is referred
to as ``specific frequency of cluster GCs'' and represented as
$S_{\rm N, CL}$. 
The error bars
in each $M_{\rm h}$  bin are estimated as $S_{\rm N,CL}/\sqrt{2(N_{i}-1)}$,
where $N_{i}$ is the total number of particles in each  bin.
The small dots represent $S_{\rm N, CL}$ 
of all group-scale  and cluster-scale  halos.
Note that dispersion of $S_{\rm N, CL}$ appears to be larger
in groups with smaller masses. 
}
\label{Figure. 13}
\end{figure}

\begin{figure*}
\psfig{file=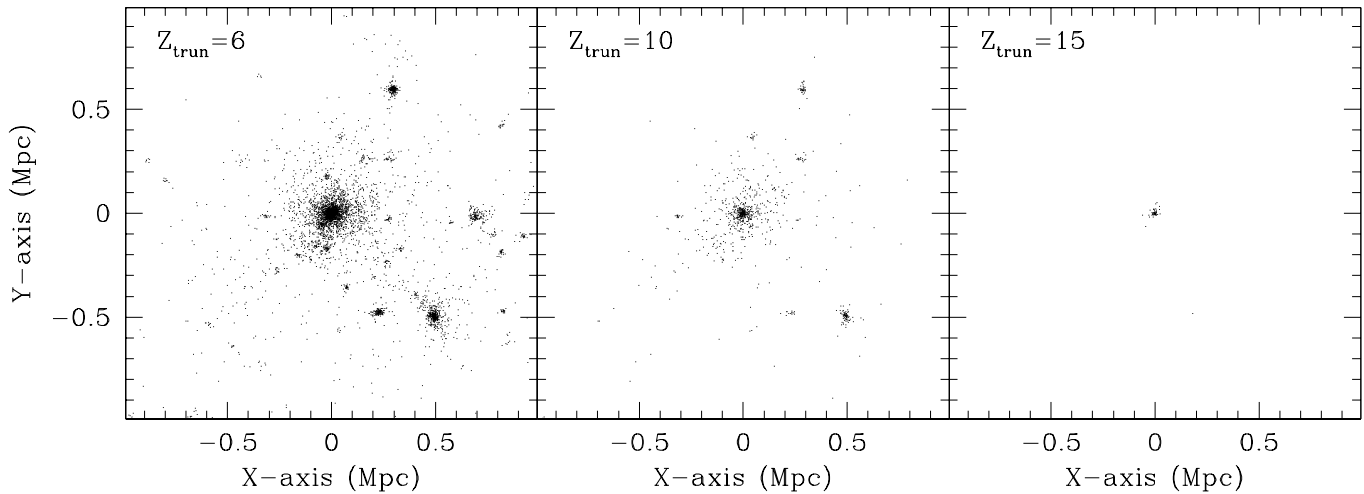,width=18.cm}
\caption{
The distributions of all GCs in the cluster CL1 for the models
with $z_{\rm trun}=6$ (left), 10 (middle), and 15 (right).
Note that the models with larger $z_{\rm trun}$ show more compact
GC distributions.
}
\label{Figure. 14}
\end{figure*}

\subsubsection{$z_{\rm trun}$}

As shown in Fig. 14, the GC distributions are quite different between
models with different $z_{\rm trun}$. The model with $z_{\rm trun}=6$ 
has a wider distribution and a larger number of substructures in the GC
distribution whereas the model with $z_{\rm trun}=15$ has a very compact
distribution with little substructures. 
The surface number densities of ICGCs in models with higher  $z_{\rm trun}$
can be lower, which implies that the ICGC distributions can give
some constraints on $z_{\rm trun}$.
The derived tendency that 
the models with higher $z_{\rm trun}$ have more compact distributions
and less substructures is discussed later in \S 4.4 in the context
of reionization influence of GC formation in clusters of galaxies.

\subsection{GNs}

The number ratio of GNs to GCs  ($f_{\rm gn}$) ranges from 0.08
to 0.14 for the clusters (CL1 - CL14) and $f_{\rm gn}$
does not depend on $M_{\rm CL}$ (almost constant at $\sim 0.1$).
The derived $f_{\rm gn}$ means  that 
the possible  GN-GC number ratio to be observed in clusters
range from $2.3 \times 10^{-4}$ to $4.1 \times 10^{-4}$
(See \S 2.2 for the conversion from the simulated value to
the observable one).
The cluster CL1 with $N_{\rm GC}=6868$
can have {\it observable} $760$ GCs at $z=0$ that were previously nuclei of
smaller galaxy-scale halos at $z=6$. 
Owing to the almost constant value of $f_{\rm gn}$ 
for different  $M_{\rm CL}$,
more massive clusters are more likely to have a larger number 
of GCs that were galactic nuclei of galaxy-scale halos  at $z=6$.

We do not find any remarkable differences in radial distributions
and kinematics within clusters between 
GNs and GCs when we separately investigate these
properties of GNs and GCs. However we find that the distributions
of GCs at $z=0$ that were formed from GNs of  massive galaxy-scale
halos with $M_{\rm h} \approx 10^{11} {\rm M}_{\odot}$ have
different distributions from other GCs in clusters.
For example,
the half-number radius ($R_{50}$) of GCs formed from GNs of massive halos
with $M_{\rm h} = 2 \times  10^{11} {\rm M}_{\odot}$ 
at $z=6$ is
0.41 Mpc for the CL1 cluster and 0.026 Mpc for the CL14 one. 
These values of $R_{50}$ are significantly smaller than
those of GCs in these clusters: $R_{50}$=0.76 Mpc for GCs of the CL1
and 0.34 Mpc for GCs of the CL14. This is due to the fact that the central
regions of clusters can be formed from more massive building blocks
(that are formed earlier) of the clusters.
Physical implications of these results are discussed later in \S 4.3.

\subsection{IGGCs}

The present simulation 
do not allow
us to make robust conclusions on the physical properties of
GCs within group-scale halos 
($M_{\rm h} \approx 10^{13} {\rm M}_{\odot}$),
because of the smaller GC number per a group 
(less than 100 GCs for virialized halos with
$M_{\rm h} \le 10^{13} {\rm M}_{\odot}$).
However,
it is still  useful to provide some qualitative results 
on the existence of intragroup GCs (IGGCs),
given the fact that extensive survey to detect bright
ICGCs or UCDs in several groups of galaxies 
are ongoing by using 2dF spectrograph
(e.g., Drinkwater et al. 2005; Kilborn et al. 2005).
Fig. 15 demonstrates that ICGCs can be formed in groups 
and the number fraction of IGGCs among all GCs ($f_{\rm IGGCs}$)
for this model is similar to $f_{\rm ICGCs}$ derived for clusters.
($\sim 0.3$). 
The results of GC properties in groups of galaxies 
(e.g., dependences of $f_{\rm IGGCs}$ on $M_{\rm h}$) will
be presented in our forthcoming papers.

\begin{figure}
\psfig{file=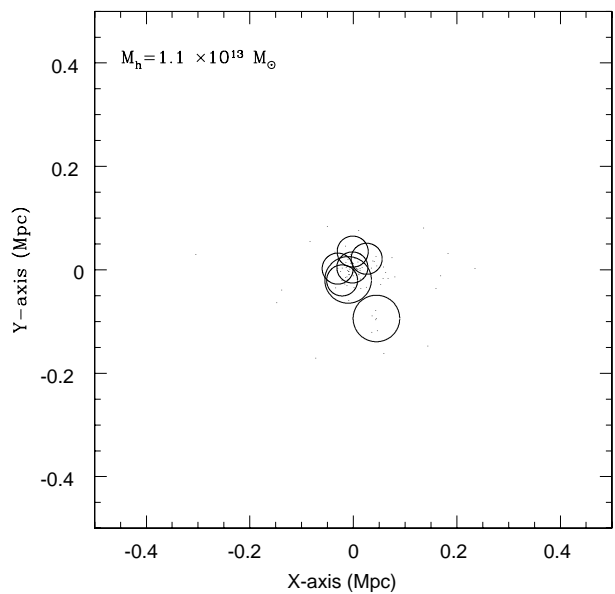,width=8.cm}
\caption{
The same as Fig. 2 but for the group-scale halo with
$M_{\rm h}=1.1 \times 10^{13} {\rm M}_{\odot}$.
}
\label{Figure. 15}
\end{figure}

\subsection{INGCs}

Our previous paper (YB) has already described the results
of INGC formation in detail.
However it would be important for readers to compare the number
fraction of INGCs with that of ICGCs derived in this paper.
Accordingly we here briefly summarize the results (YB)
as follows.

(i) INGCs comprise about 1\% of all GCs formed at $z>6$ in galaxy-scale
halos. These INGCs are formed as a result of tidal stripping
of GCs from subhalos during hierarchical structure
formation through interaction and merging of subhalos between
$z=0$ and $z=6$.

(ii) INGCs at $z=0$ were previously GCs located in the outer
parts of lower mass subhalos, which are more susceptible to
tidal stripping and destruction during hierarchical structure formation.

(iii) The number fraction of INGCs ($f_{\rm INGC}$)
is larger for $R_{\rm tr,gc}$ and this dependence
can be seen both in the model with initial GC density profile
similar to the NFW one and that with the power-law one with
the slope of $-3.5$ (i.e., similar to the observed one of
the Galactic GC system).For a reasonable set of parameters
(e.g., $R_{\rm tr,gc}$ and the FoF linking length),
$f_{\rm INGC}$ can range
from an order of 0.1\% to 1\%.

\begin{figure*}
\psfig{file=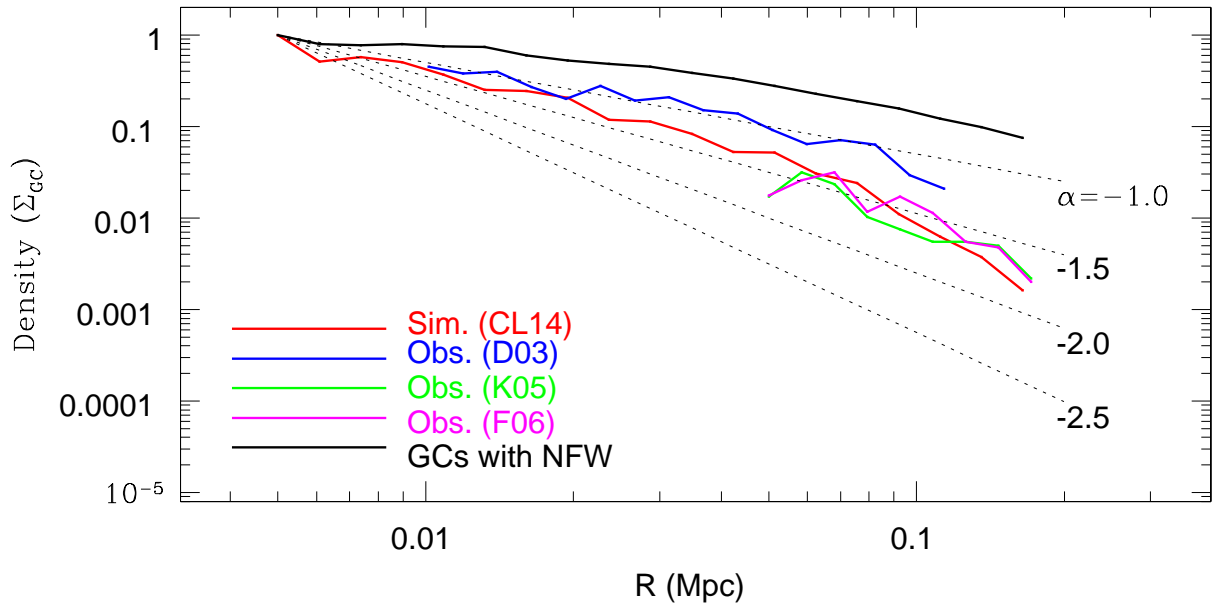,width=16.cm}
\caption{
Projected radial density profiles (normalized) 
for all GCs in the model CL1 (red), 
blue (thus possibly old) 
blue GCs around NGC 1399 and ICGC candidates
in the Fornax cluster (Dirsch et al. 2003; D03; blue),
62 UCDs in the cluster (Karick 2005; K05; green),
and 92 UCD/ICGC candidates in the cluster (Firth et al 200; F06; magenta).
For comparison,  the density profile of $10^5$ GCs with 
the ``NFW''  profile 
(model 15 in Navarro et al. 1996) in the cluster is plotted by a black line.
In order to more clearly see the slopes of the profiles,
the observed profiles (D03, K05, and F06) are vertically
shifted in a somewhat arbitrary fashion.
Thin dotted lines represent power-law slopes ($\alpha$)
of $\alpha$ = $-2.5$, $-2.0$, $-1.5$, and $-1.0$.
}
\label{Figure. 16}
\end{figure*}

\section{Discussion}

\subsection{Comparison with observations}

Recent observational studies 
based on wide-field imaging of GC systems (GCSs) 
have revealed structural properties of GCs not only
around giant galaxies (e.g.,  Rhode \& Zepf 2004)
but also across the central regions of clusters of
galaxies (e.g., Dirsch et al. 2003; Bassino et al. 2006;
Tamura et al. 2006; West et al. 2006).  
Furthermore recent observations using Two-Degree Field (2dF)
400-fiber spectrograph on the Anglo-Australian Telescope 
have discovered not only UCDs but also possible candidates
of ICGCs (Drinkwater et al. 2003; Firth et al. 2006).
Since these observations cover the GC density distributions
at most for the central $200-300$ kpc of a cluster of
galaxies (e.g., Bassino et al. 2006),
the simulated GC distributions over the entire regions
in clusters of galaxies (i.e., Mpc-scale) can not be
compared with the observations in a fully self-consistent
manner. However, it would be important to investigate
how well the simulated GC profiles can match with
the observed central GC profiles in clusters of galaxies,  
because this investigation may well  enable us to assess the viability
of the present formation model of ICGCs.

Figure 16 shows the comparison between the observed
density profiles of GCs  around NGC 1399 (Dirsch et al. 2003; D03),
UCDs (Karick 2005; K05), and ICGC/UCD
candidates  (Firth et al. 2006; F06)
in the central region ($<200$kpc) of the Fornax cluster
and the simulated one 
for all GCs including ICGCs and galactic GCs
in the model CL14 with the total
mass similar to that of the Fornax cluster.
For comparison,  the radial profile of GCs that have the NFW density
profile (model 15 in NFW)
reasonable for the Fornax cluster is also shown in this figure.
It is clear from this figure that (1) both simulated and observed
GC profiles are significantly steeper than that of the NFW profile,
(2) the profile of GCs in D03 is slightly shallower than that 
of UCDs in K05 (and ICGC/UCDs in F06),
and (3) the simulated profile can be overall consistent with 
the results of K05 and P06 and with that of D03 in the inner part
of the Fornax cluster.

The derived steeper profiles, which can be seen in models
with different $M_{\rm CL}$, can be due to the truncation
of GC formation at high $z$ 
modeled in the present study.
Because of the lack of extensive observational studies
of GC density profiles in variously different clusters
of galaxies,  it still remains unclear whether
such steeper density profiles of GCs are universal
rather than exceptional for the Fornax cluster.
Previous theoretical and numerical models of GC formation
with the truncation of GC formation at high $z$ 
were suggested to be more consistent with observed
GC properties, such as the color
bimodality of galactic GCs and the radial density
profile of the Galactic GCS (e.g, Beasley et al. 2002;
Santos 2003; Bekki 2005; Moore et al. 2005).
We accordingly suggest that future observations on
GC density profiles in different clusters are
quite important for proving the truncation of GC formation
at high $z$ by some unknown cosmological processes 
(e.g., reionization).

\subsection{Metallicity distribution function of ICGCs
}

Recent theoretical studies based on semi-analytic models
(e.g., Beasley et al. 2002) and on high-resolution
cosmological  simulations (e.g., Rhode et al. 2005; Bekki et al. 2006)
have started providing some  theoretical predictions 
on GCS properties (e.g., relations between physical properties
of GCSs and those of their host galaxies)
in a hierarchical clustering scenario
of galaxy formation (See Brodie \& Strader 2006
for a recent review). However these studies have not yet
provided useful predictions on ICGC physical properties 
(e.g., metallicity distributions functions that are referred to as
MDFs in this paper) that can be  compared with
previous/ongoing observations of ICGCs (e.g., Hilker 2002;
West et al. 2006). In particular,  the observed color distributions
of ICGCs, which contain information on metallicities of ICGC
and thus on chemical evolution histories of ICGCs and their
defunct host galaxies, can be important to be compared with 
the present results.

The methods to investigate MDFs of ICGCs and GCs within galay-scale
halos in clusters of galaxies are described as follows.
Firstly we allocate metallicities ($Z_{\rm gc}$) 
to GCs within a virialized halo with the total mass ($M_{\rm h}$) and
the  baryonic  one ($M_{\rm b}$) at $z=z_{\rm trun}$ by assuming
a reasonable relation between $Z_{\rm gc}$ and $M_{\rm b}$
(and $M_{\rm b}/M_{\rm h}$).
We here adopt the observationally suggested  $Z_{\rm gc}-M_{\rm b}$ relation
by Peng et al. (2006) that is described as $Z_{\rm gc} \propto
{M_{\rm b}}^{0.5} $ (for a constant stellar mass-to-light-ratio).
Since we do not have observational data sets that can 
give strong constraints on the $Z_{\rm gc}-M_{\rm b}$ 
relation {\it at high redshifts},
we consider that the adopt relation 
is a reasonable first step for better understanding MDFs of GCs 
in the CDM model.
Secondly, we follow the dynamical evolution of GCs with
different $Z_{\rm gc}$ until $z=0$, and then investigate
$Z_{\rm gc}$ of ICGCs and galactic GCs in halos identified
as clusters of galaxies. 

Fig. 17 shows that the MDF of ICGCs in the model CL1 
is significantly different from that of GCs within galaxy-scale
halos (GGCs)  in the cluster  in the sense that
the MDF of ICGCs   has a much higher fraction of metal-poor
([Fe/H] $<-1.6$) GCs than that of GGCs.
The mean [Fe/H] for ICGCs, GGCs, and all GCs are $-1.45$,
$-1.12$, and $-1.21$, respectively, in this model.
The ICGC MDF does not show  the  remarkable  bimodality
that is observationally suggested to be common in 
bright galaxies (e.g., Brodie \& Strader 2006).
Interestingly,  the MDF of GGCs shows relatively clear bimodality
in this model without no GC formation via 
star formation at $z<6$.
The reason for the high fraction of metal-poor GCs
in the ICGC population is that ICGCs originate from
lower mass galaxy-scale halos that can be more easily
destroyed by the cluster tidal to disperse their GCs
into intracluster regions.

By using relations between colors and metallicity (e.g., Barmby et al. 2000),
we can convert MDFs of ICGCs into their color distributions
that can  be compared with ongoing and future observations. 
The derived peak of [Fe/H] $\sim -1.6$ corresponds to
$V-I \sim 0.9$, which is roughly consistent with the observed
peak value of ICGC candidates
for the ``region 2'' in the Centaurus cluster of galaxies
(Hilker 2002). Because of the linear dependence of $V-I$ on
[Fe/H] in the adopted $V-I$-[Fe/H] relation (Barmby et al. 2002),
the simulated MDF of ICGCs can not be consistent with
the observed MDF similar to a Gaussian profile for
ICGC candidates in the Centaurus cluster (Hilker 2002).
This apparent inconsistency is due largely to the derived
large fraction of metal-poor GCs in the simulated ICGC 
populations.
It should be here noted that the observed MDF also
shows no remarkable bimodality: It however remain unclear
whether such an unimodal distribution is an unique character
in MDFs of ICGCs. 

Although the above apparent inconsistency between observations
and simulations in MDFs (or color distributions) of ICGCs
may well result either from 
numerical resolutions of the present study  (i.e., not enough high-resolution
to properly treat low-mass halos susceptible to tidal
destruction) or from some uncertainties in the modeling
of initial $Z_{\rm gc}-M_{\rm b}$ relations, 
this inconsistency 
most likely suggests that
formation of GCs at high $z$ in low-mass halos (or galaxies),
where most low-$Z_{\rm gc}$ ICGC originate from,
needs to be suppressed in order 
to explain  more self-consistently  observations.
It is however unclear what physical mechanisms are responsible
for such preferential suppression of GC formation in
low-mass halos.
We thus plan to investigate whether
theoretical  models with the preferential suppression of GC formation
can better explain the observed MDFs of ICGCs 
by using more sophisticated and higher-resolution 
numerical simulations.

In the present models, all ICGCs are initially within galaxy-scale
halos at very high redshifts ($z>6$) and later can become ICGCs owing to  tidal
stripping of GCs from the halos during slow hierarchical growth of
clusters of galaxies via halo merging (See Fig. 3).
GCs within the halos therefore can have enough time to be dynamically
influenced by the tidal fields of the halos 
and galaxies embedded in the halos  before they are stripped
to form ICGCs.
Recent dynamical studies of GCSs in galaxies
have demonstrated that 
tidal fields of galaxies are highly likely
to destroy preferentially low-mass GCs so that luminosity functions (LFs)
of GCs can have log-normal shapes rather than
power-law ones (e.g., Fall \& Zhang 2001).
If this result can be applied to GCs in galaxy-scale halos that
are building blocks of clusters of galaxies,
LFs of ICGCs are highly likely to  have log-normal shapes.
We thus suggest that the observed LFs of ICGCs can provide
some clues for a question as to  whether ICGCs were initially 
within galaxy-scale
halos at high $z$.

\begin{figure}
\psfig{file=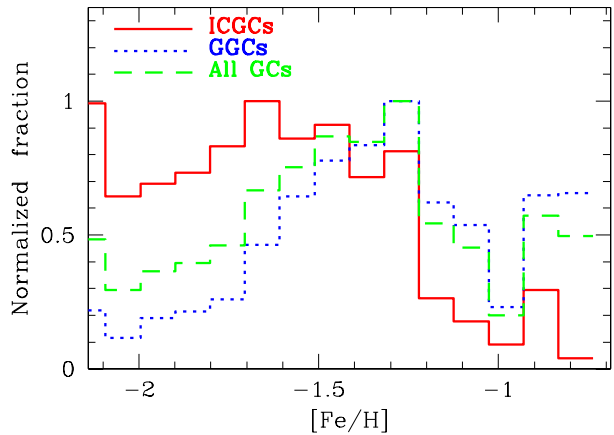,width=8.cm}
\caption{
The simulated metallicity distribution functions (MDFs, normalized)
for ICGCs (red and solid), galactic GCs (GGCs,  blue and dotted),
and all GCs (green and dashed) in the model CL1
for $M_{\rm h}/M_{\rm b}=20$ and $Z_{\rm gc}=0.0029$ 
(or [Fe/H]$=-0.84$) in
galaxy-scale halos with 
$M_{\rm b}= 6 \times 10^{10} {\rm M}_{\odot}$.
Note that the ICGC MDF does not show remarkable bimodality.
}
\label{Figure. 17}
\end{figure}

\subsection{Origin of high $S_{\rm N}$ of Es in clusters}

It is well known that giant Es in the central regions 
of some clusters of galaxies (e.g.,  M87 in the Virgo and NGC 1399 in
the Fornax) have much  higher  $S_{\rm N}$  ($>10$) in comparison
with field Es (e.g., Harris 1991; West et al. 1995; Forbes et al. 1997).
Forbes et al. (1997) found that Es with high $S_{\rm N}$ have
higher number ratios of metal-poor GCs (MPCs) to metal-rich ones (MRCs)
and accordingly suggested that the secondary formation of MRCs 
does not explain clearly the origin of very high $S_{\rm N}$
in some cluster Es.
West (1993) proposed that the observed high $S_{\rm N}$ of central Es
in rich clusters can be naturally explained if GCs can form more
efficiently in rare high-density peaks of primordial matter distributions.
This ``biased GC formation scenario'' was suggested to be consistent
with the observed dependence of  $S_{\rm N}$ on environments (e.g.,
mean galaxy density) and galaxy types (West 1993).

The present study has shown that a large fraction (up to 36 \%) of
old GCs formed before $z=6$ can be within the central 50 kpc of clusters,
though the GC number can be quite different between clusters with different
$M_{\rm cl}$.  
For example, the model CL1 shows $N_{\rm GC,0.05}$ of 1098, 
which means that about 16 \% of all GC 
particles are located within the central 50 kpc of the cluster. 
If the cluster has a central giant E with the luminosity
similar to M87 ($\approx -22.7$ mag in $V$-band; West et al. 1995),
this result can be interpreted as the cluster having the
central E with 15350 GCs and $S_{\rm N}$  of $\approx 13.4$
(See the Table 1 in West et al. 1995).
As shown in the present simulations,
the origin of a large number of GCs confined in the central region
of a cluster is closely associated with the fact
that the cluster central region is formed by hierarchical merging
of halos which were virialized at very high redshift ($z >6$)  
and thus contained  GCs.
Thus the present simulations imply that the origin of 
very high $S_{\rm N}$ of giant Es in the central regions of
clusters can be understood in terms of the growth processes 
of the clusters.

The present study has also shown that  
$N_{\rm GC,0.05}$ is different by a factor of 13 between
different clusters (for $M_{\rm CL} > 10^{14} {\rm M}_{\odot}$). 
This result may well 
provide physical basis for the West et al's (1995) scenario
that the significant difference in $S_{\rm N}$  between central giant Es
in clusters is due to the difference in number of GCs trapped by
cluster potential (rather than galaxy one) between different clusters.
We suggest that the difference in the GC number can result from
the differences in the growth histories of clusters 
via hierarchical merging:If a cluster is formed from a larger 
number of smaller halos that were virialized before $z_{\rm trun}$
(thus could have old GCs),
it shows a higher $S_{\rm N}$ in its central giant E(s).
The derived weak dependence of $N_{\rm GC,0.05}$ on $M_{\rm CL}$
will be able to be tested against 
observational results based on wide-field
photometric studies of GCs in clusters (e.g., Dirsch et al. 2003)

\subsection{Formation of flatter GC density profiles in central giant Es
in clusters}

It has long been known that the radial density profiles of
GCSs (${\Sigma}_{\rm GC} (R)$) in elliptical galaxies
vary with the total luminosities of their host galaxies
(e.g., Ashman \& Zepf 1998).
Observations show that if ${\Sigma}_{\rm GC} (R)$ is described as 
${\Sigma}_{\rm GC} (R)  \propto R^{\alpha}$,
${\alpha}$ range from $\sim -2.5$ (for lower luminosity Es)
to $\sim -1$ (for higher).
The key point here is that
very bright cluster Es with $M_{\rm V} <-22$ mag (e.g., M87)
have very flat distributions with ${\alpha} > -1.5$.
Previous theoretical studies have pointed out that destruction
of GCs is a key physical process that controls the radial profiles
of GCSs (e.g., Baumgardt 1998; Fall \& Zhang 2001; Vesperini et al. 2003),
and some of these suggested that the very flat GC profiles
for giants Es in clusters (e.g., M87)
is due to GC destruction processes (e.g., Vesperini et al. 2003).

An alternative scenario has been proposed by
Bekki \& Forbes (2006), in which 
the radial density profiles of GCSs
in Es become progressively flatter as the galaxies
experience more major merger events.
The expected effect of 
destruction mechanisms on the GC luminosity function
with galactocentric radius was not found in the detailed studied
by Harris et al. (1998) of M87, which implies that the former
scenario is not so promising.
However it remains unclear which of the above two scenarios
is more realistic and convincing.

The present study has shown that 
the radial profiles of  GCs in the central regions of clusters
can become quite flat ($ -1.5 < \alpha  < -1.0$) 
{\it without any destruction mechanisms of GCs around Es}.
This result may well lend support to 
the latter scenario of the above two
(i.e., Bekki \& Forbes 2005),
though the present model does not consider (1) the formation
of Es  in the central regions of clusters
and (2) the contribution of MRCs (which could be less than 30\%
in number of Es with high $S_{\rm N}$) to the GC density  profiles.

We accordingly  suggest that  the observed flat GC distributions
of cluster giant Es are closely associated with the formation 
processes of the cluster central regions (thus central giant Es) 
via hierarchical merging.
Since the merging histories of clusters are different in
different clusters,  the observed appreciable differences
in $\alpha$ between cluster Es (Ashman \& Zepf 1998) 
could be due to the different merging histories of clusters.

\subsection{UCDs as nuclei of oldest galaxies}

One of the promising scenarios of UCD formation
is the ``galaxy threshing'' one (Bekki et al. 2001, 2003) in which
dE,Ns can be transformed into UCDs owing to complete tidal
stripping of outer stellar envelopes of dE,Ns by strong
tidal field of clusters of galaxies.
In this scenario, dE,Ns that are transformed  into UCDs
are required to 
have higher eccentricity and pericenter distances of their orbits.
This requirement naturally explain the observational result
that  bright UCDs in the Fornax cluster are 
confined within 
the central 200 kpc of the (Drinkwater et al. 2000; Bekki et al. 2003).
In order for the threshing scenario to be consistent
with the observed number of UCDs, 
clusters of galaxies should contain a large number of dE,Ns
that could be progenitor galaxies of UCDs. 
Observations revealed that (1) dEs located near the center
of the Virgo cluster are mostly nucleated
and (2) only a fraction of dEs located outside the cluster core
radius are non-nucleated (Binggeli \& Cameron 1991).
The threshing scenario alone has not yet explained why
dEs are mostly nucleated in the central regions of clusters of galaxies.

The present study has shown that 
(1) cluster GCs at $z=0$ originating from GNs of
massive galaxy-scale halos ($M_{\rm h} \approx 10^{11} {\rm M}_{\odot}$)
at high redshifts
have compact distributions and 
(2) GCs formed in  halos virialized 
at very high redshift ($z>10$) can be located preferentially in the
central regions of clusters at $z=0$ (See Fig. 14).
Although cluster GC distributions can not be directly compared with
UCD ones,
these results may well imply that
{\it if UCDs were nuclei of dE,Ns formed in the massive halos 
virialized 
at very high redshifts and later became cluster member galaxies,
then they 
can have very compact distributions in clusters at $z=0$}:
The origin of the observed 
compact distributions of UCDs (and dE,Ns) can be  due to
the earlier formation of massive dE,Ns in comparisons with non-nucleated dEs.
This somewhat speculative explanation needs to answer 
the question as to why massive halos virialized at very high redshift
can develop stellar nuclei in their central regions
and consequently become nucleated galaxies.
We suggest that higher mass densities in halos virialized at
earlier epoch can be responsible for more efficient nucleus formation
either by merging of star clusters 
or by some dissipative mechanism in galactic centers
(Oh \& Lin 2000; Bekki et al. 2004; Bekki et al. 2006).

\subsection{Reionization and GC properties in clusters}

Recently several authors have discussed (1) whether 
reionization can trigger 
or suppress the formation of globular clusters (e.g., Cen 2001;
Santos 2003)
and (2) what observational properties of GCSs in galaxies
have fossil information of the reionization influence
on GC formation (Bekki 2005).
These studies proposed that $S_{\rm N}$ of MPCs in giant Es,
the color bimodality of GCSs,
and the projected radial density profile of GCSs
can be influenced by reionization, {\it if
reionization can suppress GC formation in subgalactic clumps
(i.e., progenitors of dwarf galaxies).}
These previous studies however did not discuss at all
whether physical properties of GCs (i.e., radial distributions
of ICGCs) in clusters of galaxies can have fossil records of
reionization influence on GC formation in clusters.

The present study has shown that the spatial distributions of
GCs in clusters  depend strongly on $z_{\rm trun}$ in the sense that
the cluster models with higher  $z_{\rm trun}$ show more compact
GC distribution. 
This result implies that if $z_{\rm trun}$ is physically related 
to the epoch of the completion of reionization ($z_{\rm reion}$), after which
GCs formation is strongly suppressed,
the GC density distributions  in clusters can have some valuable
information on $z_{\rm reion}$.
Very deep  `all-object'  spectroscopic surveys centered on
some clusters are required for identifying cluster member GCs and 
thereby investigating radial density profiles of GCs.
This type of observations might well be formidable tasks for
8m-telescopes with multi-object spectrograph
and wide-field imaging facilities.
Systematic observational studies of radial density profiles
of cluster GCs  by
such telescopes will enable us to discuss differences in reionization
influences on GC formation between different clusters 
and thus to understand better the origin of cluster GCs.

\section{Conclusions and summary}

We have investigated radial density profiles and kinematics of 
GCs located in clusters with the total masses ($M_{\rm CL}$)
larger than $10^{14} {\rm M}_{\odot}$ based on high-resolution 
cosmological N-body simulations with a formation model of old GCs.
Although YB has already described some preliminary results
for only one cluster model,
this paper first discussed the dependences of physical properties
of ICGCs on $M_{\rm CL}$ based on the results on 14 cluster models. 
We summarize our principle result as follows.

(1) GCs located outside the tidal radii of galaxy-scale halos
in clusters of galaxies are formed owing to tidal stripping
of GCs initially within smaller halos 
during hierarchical growth of clusters via halo merging.
These intracluster GCs (ICGCs) comprise 20-40 \% of all GCs
in clusters with 
$1.0 \times 10^{14} {\rm M}_{\odot} \le M_{\rm CL} \le 
6.5 \times 10^{14} {\rm M}_{\odot}$ and the number fraction
($f_{\rm ICGC}$) does not depend on $M_{\rm CL}$ for the above
cluster mass range.

(2) The projected radial density profiles 
(${\Sigma}_{\rm GC}$) of all GCs
in clusters with different $M_{\rm CL}$
can be diverse. If ${\Sigma}_{\rm GC} (R) \propto R^{\alpha}$,
$\alpha$ ranges from $\approx -1.5$ to $\approx -2.5$
 for GCs in clusters with
the above mass range. 
${\Sigma}_{\rm GC}$ is more likely to be significantly flatter
in ICGCs than in GCs  located in virialized galaxy-scale halos
(i.e., GGCs) in clusters.
There are no remarkable difference in kinematical properties
(e.g., $V_{\rm r}$ distributions) between GGCs and ICGCs.

(3) Two-dimensional distributions of ICGCs are inhomogeneous,
asymmetric, and somewhat elongated,
in particular, in the outer parts of cluster for most models. 
The simulated inhomogeneous distributions of ICGCs suggest
that observations based on wide-field imaging of GCs in
clusters are doubtlessly worthwhile to understand
real radial distributions of ICGCs. 

(4) Although total number of GCs within the central 0.05 Mpc ($N_{\rm GC,0.05}$)
and 0.2 Mpc ($N_{\rm GC,0.2}$) are diverse in different clusters,
they can depend
weakly on $M_{\rm CL}$ in such a way that both $N_{\rm GC,0.05}$
and $N_{\rm GC,0.2}$ are likely to be
larger for clusters with larger $M_{\rm CL}$.
This result implies that central giant ellipticals (Es) in more massive
clusters are likely to show a larger number of  old, metal-poor
GCs compared with those  in less massive clusters.
This result also provides a new clue as to why  some central Es in clusters
show high $S_{\rm N}$. 

(5) ${\Sigma}_{\rm GC}$ of all GCs in the central 200 kpc in
clusters is also
diverse with the slopes $\alpha$ ranging from 
$\approx -1.0$ to $\approx -2.0$.
The derived flat distributions of GCs in the central region of clusters
suggest that the origin of the observed flat distributions of
GCs around central giant Es in clusters (e.g., M87 in the Virgo)
can be understood in terms of hierarchical growth processes of
the cluster central regions via merging of halos with old GCs.

(6) The distributions of GCs depends on $z_{\rm trun}$,
after which GC formation is assumed to be truncated by some physical
processes at high redshifts.
For example, the projected radial GC  distributions
are more compact and steeper for the cluster models 
with higher $z_{\rm trun}$.
ICGC fraction can be smaller for the models with higher $z_{\rm trun}$.
This result suggests that if the suppression mechanism of GC formation
is due to reionization (Santos 2003;
Bekki 2005; Moore et al. 2005), the GC distributions in clusters can provide some
useful constrains on the epoch of reionization. 

(7) If ultra-compact dwarf galaxies (UCDs),
which can be regarded as very massive INGCs,
originate from  nuclei of nucleated dwarfs (dE,Ns) that were virialized
very high $z$ ($z>10$), the origin of 
the observed compact spatial distributions
can be more clearly understood.

(8) About 1\% of all GCs formed before $z>6$ are not located
within any virialized galaxy-scale, group-scale, and  cluster-scale halos
and thus can be regarded as INGCs that are freely drifting in intergroup
and intercluster regions. 
This result is already discussed extensively  by YB.
Physical properties
(e.g., metallicity distributions and luminosity function)
of these intergroup and intercluster
GCs can be  quite different from those of GCs within galaxies.

(9) The mean metallicity of ICGCs in a cluster can be smaller than that
of GCs within the cluster member galaxy-scale halos by $\sim 0.3$
in [Fe/H].
Metallicity distribution functions (MDFs)  of ICGCs show 
peak values around [Fe/H] $\sim -1.6$ with a higher fraction
of metal-poor GCs (with [Fe/H] $<-1.6$). 
The MDFs do not have 
remarkable  bimodality that is observationally suggested
to be common features of MDFs in bright galaxies.

(10) Radial density profiles of ICGCs and galactic GCs in 
clusters of galaxies can be significantly steeper than
those of the background dark matter halos in the central
regions ($R<200$kpc) of the clusters
with different  $M_{\rm CL}$. 
This is due mainly to the truncation of GC formation at high $z$ ($>6$)
in the simulations.
The simulated GC profiles can be overall consistent with
latest observations on radial density profiles of GCs and UCDs 
in the Fornax cluster of galaxies.

\section*{Acknowledgments}
We are  grateful to the referee for valuable comments,
which contribute to improve the present paper.
We are also grateful to  Arna Karick and Peter Firth for their sending
observational data sets of intracluster GCs and ultra-compact dwarfs. 
KB  acknowledges the financial support of the Australian Research 
Council throughout the course of this work.
The numerical simulations reported here were carried out on
Fujitsu-made vector parallel processors VPP5000
kindly made available by the Astronomical Data Analysis
Center (ADAC) at National Astronomical Observatory of Japan (NAOJ)
for our  research project why36b.
H.Y. acknowledges the support of the research fellowships of the Japan
Society for the Promotion of Science for Young Scientists (17-10511).

\appendix

\section[A]{Identification of  galaxy-scale halos in clusters}

We adopted the following method in order to identify galaxy-scale halos
in a cluster-scale halo. Firstly, we investigate 
the mass density (${\rho}_{\rm i}$)  
around each ($i$-th) particle in a cluster-scale halo
and thereby sort out ${\rho}_{\rm i}$ for all particles in the cluster
(i.e., the particle with highest ${\rho}_{\rm i}$ 
is the first and that with
the lowest is the last in  order).
Secondly, by assuming that the particle with the highest  ${\rho}_{\rm i}$
can be the ``nuclear particle'' which
is located in the central region of a subhalo, we identify this nuclear
particle. Thirdly, we look for the tidal radius $R_{\rm T}$ (which 
is described in the Appendix B in detail) 
for this subhalo, and particles within $R_{\rm T}$
are regarded as being the member particles of this halo.
Fourthly, we assume that the particle 
with the second highest ${\rho}_{\rm i}$ 
is the nuclear particle of an another subhalo and thereby look for the
member particles of the halo. This process is repeated for 3rd, 4th, etc.
until every galaxy-scale halos are found.
The GC and GN (``galactic nucleus'') particles that do not belong to any
subhalos are identified as ICGCs (intracluster GCs). We have confirmed
that this method enables us to correctly identify subhalos
though the above first process can be time-consuming in some cases.

\section[B]{Methods to determine  tidal radii of GCSs in galaxy-scale halos}

In order to select ICGCs from GCs in a cluster-scale halo,
we need to find GCs outside the tidal radii ($R_{\rm t}$)
of GCSs in galay-scale halos in the cluster. 
Because of tidal stripping of dark halo particles and GC ones,
the radial density profiles of
the outer parts of galaxy-scale halos can much more steeply
decrease than dynamical equilibrium models predict.
We define $R_{\rm t}$ of a  GCS as the radius beyond which
the radial density profile of the GCS (${\rho}_{\rm GC} (r)$)
drops much more sharply than the CDM model predicts. 
The universal density profiles of 
dark matter halos are predicted by extensive cosmological
simulations such as 
Navarro, Frenk, \& White (1996; hereafter NFW) and
the NFW profile is described as:
\begin{equation}
{\rho}(r)=\frac{\rho_{\rm s}}{(r/r_{\rm s})(1+r/r_{\rm s})^2},
\end{equation}
where $r$,  $\rho_{\rm s}$,  and $r_{\rm s}$ 
are the distance from the center
of the dark matter halo,  
the characteristic density, and the scale-length of the dark halo,
respectively. This NFW profile means that if dark matter 
halo density profiles are described as $r^{\gamma}$ for 
the outer parts ($r \gg  r_{\rm s}$),
${\gamma} \approx -3$. 
By assuming that ${\rho}_{\rm GC} (r) \propto r^{\beta}$
for a GCS in a galaxy-scale halo,
we first estimate the radius
where $\beta$ becomes smaller than $-5$
(which is significantly smaller than $-3$ in the outer NFW profile),
and then regard the radius as $R_{\rm T}$.

$R_{\rm T}$ determined in the above way for a
galaxy-scale halo in a cluster would not be exactly the
same as the tidal radius that can be determined from
the total mass of the cluster and the mass and
the orbital eccentricity of the galaxy-scale halo
(e.g., Faber \& Lin 1983). However, we consider that
the method described above is much more practical
and thus better,
because (1) observational studies can find $R_{\rm T}$ 
for each individual GCS in the same way as the present 
simulation does so that we can compare simulations with
observations in a fully self-consistent manner
and (2) it is currently impossible to observationally determine
the orbital eccentricity of each individual halo with
respect to the cluster center
owing to lack of proper motion data for cluster GCs.
Since this new method of determining
$R_{\rm T}$ is based solely on the positions of GCs,
we can efficiently  find ICGCs that are outside $R_{\rm T}$
of any galaxy-scale halos in a cluster.

Based on the above method to determine $R_{\rm T}$,
we investigate total number of GCs within each individual halo
in each cluster model (CL1-14). 
Fig. B1 shows the ``GC number ($N_{\rm GC}$) function''  
(like galaxy luminosity function)
for  galaxy-scale halos
in the cluster CL1 at $z=0$
and for all galaxy-scale halos at $z=6$.
This  $N_{\rm GC}$ function describes how much fraction
of galaxy-scale halos contain a given number of GCs. 
As shown in Fig. B1, the $N_{\rm GC}$ function depends on 
redshifts  in the sense that a larger fraction of galaxy-scale
halos can have more than 100 GCs within  their tidal radii 
at $z=0$ as a result of hierarchical merging.
Some halos in Fig. B1 show $\sim 1000$ GCs within their
tidal radii, and these halos are the central giant ones
in the cluster CL1. 

\begin{figure}
\psfig{file=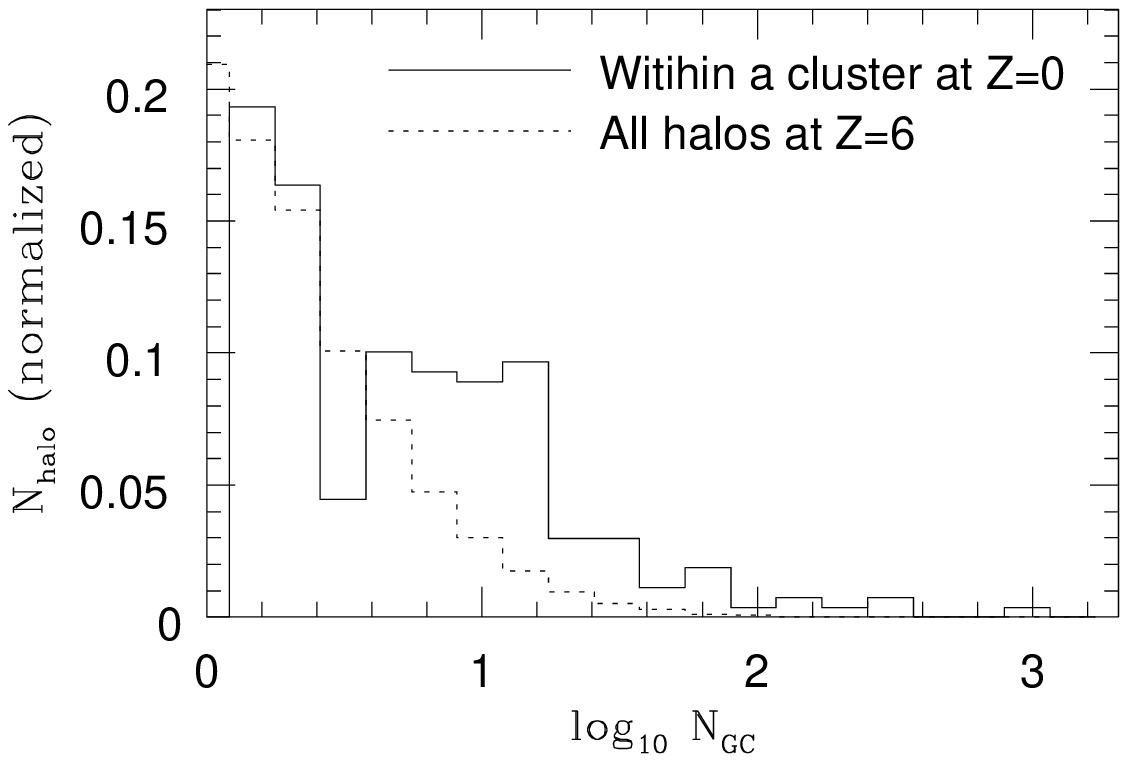,width=8.cm}
\caption{
The GC number ($N_{\rm GC}$) function for the cluster CL1 at $z=0$ (solid)
and all halos at $z=6$ (dotted).  
A clear evolution of the $ N_{\rm GC}$ function 
can be seen between $z=6$ and 0.
}
\label{Figure. 16}
\end{figure}

\section[C]{Dependences of the results on $f_{\rm gc}$}

This parameter $f_{\rm gc}$  can set the level of statistical sampling of the
GCs in the present simulations. To show the robustness of the present results,
we investigated the models with different $f_{\rm gc}$  and the Fig. C1
shows one example of the $f_{\rm gc}$-dependences of the results.
As shown in Fig. C1,  
$f_{\rm INGC}$
(number fraction of intergalactic GCs)  does not depend on
$f_{\rm gc}$ for $f_{\rm gc}>0.05-0.1$.
Considering that the baryonic fraction of halos (thus GC mass fraction)
can be at most 0.2, the above result
suggests  that reasonable values of $f_{\rm gc}$ should be $0.1 - 0.2$
in the present simulations. 
We confirm that the present results do not depend on
$f_{\rm gc}$ for $0.1 \le f_{\rm gc} \le 0.2$.

\begin{figure}
\psfig{file=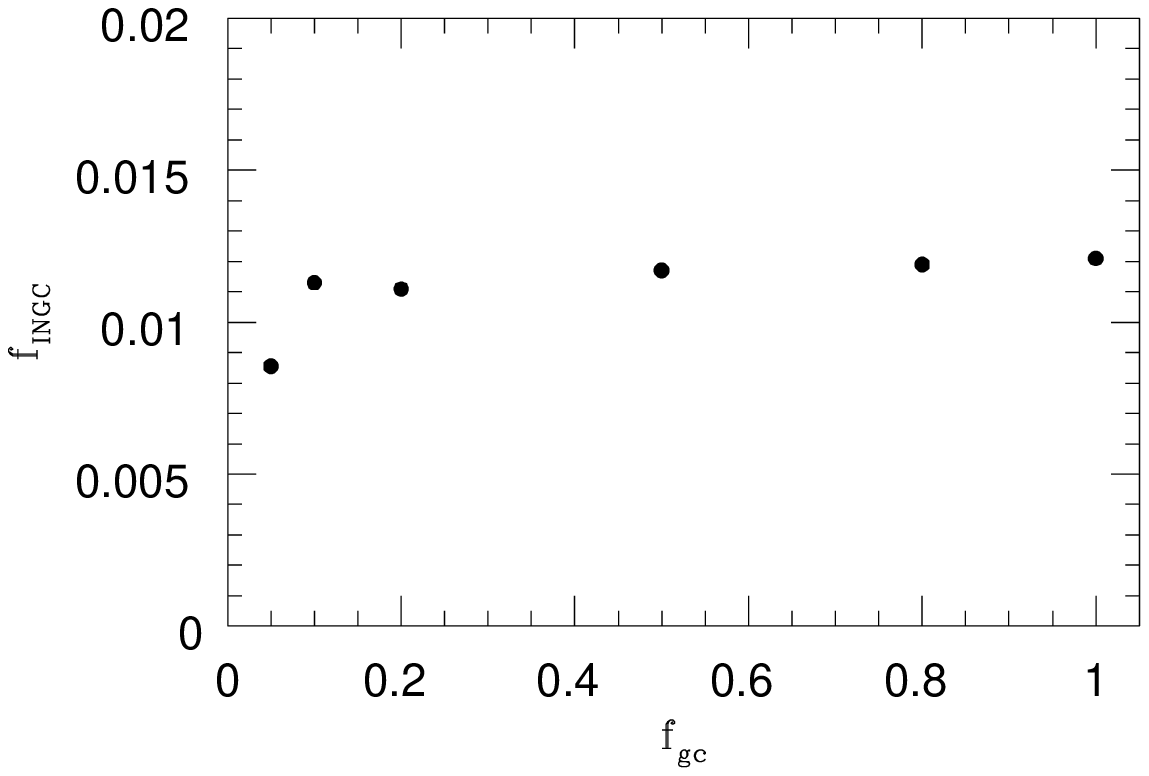,width=8.cm}
\caption{
The dependence of number fraction of INGCs ($f_{\rm INGC}$) on
$f_{\rm gc}$. Note that the results do not depend on
$f_{\rm INGC}$ for $f_{\rm INGC} \ge 0.1$.
}
\label{Figure. 19}
\end{figure}

\end{document}